\documentclass[twocolumn,aps,pra,superscriptaddress,nofootinbib]{revtex4-2}

\usepackage{graphicx}
\usepackage[dvipsnames,svgnames]{xcolor}
\usepackage{bm,bbm}
\usepackage{braket}
\usepackage{amsthm,amsmath,amssymb}
\usepackage{enumerate}
\usepackage{booktabs,multirow}
\usepackage{mathtools,bbding}
\usepackage[percent]{overpic}
\usepackage{microtype}
\usepackage{array,adjustbox,afterpage}
\usepackage[T1]{fontenc}
\usepackage[utf8]{inputenc}
\usepackage{tikz}
\usetikzlibrary{calc}
\usepackage[english]{babel}
\usepackage{newtxtext,newtxmath}
\usepackage{dsfont}

\usepackage[linesnumbered,ruled]{algorithm2e}  

\makeatletter
\renewcommand{\fnum@figure}{FIG. \thefigure} 
\makeatother
\makeatletter
\renewcommand{\fnum@table}{TABLE \thetable} 
\makeatother

\usepackage[bookmarks=false, colorlinks=true, linkcolor=blue, citecolor=purple, urlcolor=purple]{hyperref}
\usepackage{cleveref}
\usepackage[normalem]{ulem}

\Crefname{subfigures}{figure}{figures}
\Crefname{subfigures}{Figure}{Figures}

\def\bra#1{\mathinner{\langle{#1}|}}
\def\ket#1{\mathinner{|{#1}\rangle}}

\newcommand\Tstrut{\rule{0pt}{2.6ex}}         
\newcommand\Bstrut{\rule[-0.9ex]{0pt}{0pt}}   

\newcommand{\xdownarrow}[1]{{\left\downarrow\vbox to #1{}\right.\kern-\nulldelimiterspace}} 

\hypersetup{
    bookmarksnumbered=false, 
    breaklinks=true,
    unicode=false, 
    pdfstartview={FitH}, 
    pdftitle={}, 
    pdfnewwindow=true, 
}

\begin{document}


\title{Manipulating fermionic mode entanglement in the presence of superselection rules}

\author{Ömer Tırınk}
\email{omertirink@outlook.com}
\affiliation{Tofaş Science High School, 16120 Nilüfer, Bursa, T\"{u}rkiye}

\author{Gökhan Torun}
\email{gtorung@gmail.com}
\affiliation{Department of Computer Engineering, Alanya University, 07400 Alanya, Antalya, T\"{u}rkiye}

\author{Onur Pusuluk}
\thanks{corresponding author}
\email{onur.pusuluk@gmail.com}
\affiliation{Faculty of Engineering and Natural Sciences, Kadir Has University, 34083, Fatih, Istanbul, T\"{u}rkiye}

\begin{abstract}
Superselection rules (SSRs), linked to the conservation of physical quantities such as parity or particle number, impose constraints on allowable physical operations in fermionic systems. This affects the amount of extractable mode entanglement possessed in a given state and its manipulation by the so-called entanglement-free operations. Here, we present a majorization-based algorithm for the mixed state transformations of bipartite mode entanglement, where the allowed operations (i.e., resource non-generating operations), that is, local operations and classical communication, are restricted by local SSRs. We then focus on the local parity SSR and investigate the possibility to relax the constraints imposed by it through a catalyst. In particular, we show that an ancillary mode system can catalyze the change in local parity. Finally, we discuss the application of our methodology to various problems in different fields. Accordingly, we propose that it may shed new light on the activation of orbital entanglement in chemical molecules and the manipulation of multipartite entanglement or quantum discord in distinguishable quantum systems.
\end{abstract}

\maketitle


\section{Introduction}

Over the past decade, quantum resource theories (QRTs) have come to be a study of how the quantum properties could lead to an operational advantage in quantum information processing and broadly speaking, concerned with their quantification, characterization, and manipulation under physical restrictions \cite{OppenheimQRT13, ChitambarQRT19, TJP2023}. Essentially, a quantum resource theory is composed of two elements: the free states (whose set is denoted by \(\mathbb{F}\)) and the free operations (denoted by \(\mathbb{O}\)). All states which are not free are regarded as (quantum) resources. By definition, free operations are physical transformations that do not create any resources, that is, they must transform free states into free ones and allow for the resource to be manipulated but not freely created.

The picture of a hierarchy of quantum states --- a structural arrangement based on resource contents --- has been well grounded, in both theory and practice, by means of state transformations in the resource theory of bipartite entanglement \cite{Luigi2008ENT, Horodecki-QE} and coherence \cite{Baumgratz-Coherence} in distinguishable systems. Specifically, making use of the concept of majorization \cite{Bhatia, Marshall}, Nielsen's theorem \cite{Nielsen-Maj} provides the necessary and sufficient conditions for the transformations that take one given bipartite pure entangled state to another with a unit probability of success, where the allowed (i.e., free)  operations are local operations and classical communication (LOCC). Following this pioneering work \cite{Nielsen-Maj}, majorization became a crucial instrument in several studies, such as entropic uncertainty relations \cite{Augusiak_2009, Partovi2011EntMaj, Puchala2013ENT}, quantum algorithms \cite{Delgado2002MAJQA, Zhaohui2006MALG, Vallejos2021MAJ}, convertibility of resource states \cite{Vidal2001InterConv, TorunENTTR15, Du-CoherenceMaj, Huangjun17COHENTMAJ, Horodecki2018ApprMaj, TorunCOHTR18, Bosyk2019MAJTR, BosykMAJ21, CundenMAJ21}, and quantum thermodynamics \cite{Horodecki2013TMaj, GOUR2015TMajor, Renes2016SubMaj, Gour2018TMaj, SinghHOFMAJ21}.

This paper aims to provide an extension of the resource theory of bipartite entanglement to include systems consisting of indistinguishable particles, such as chemical molecules. Of course, to take steps in this direction, one can consider either particle entanglement or mode entanglement \cite{Zanardi2002ModeE, Yu2003ModeE, Vaccaro2003ENTID, Friis2013ModeE}, both of which are distinct from entanglement between distinguishable systems \cite{Luigi2008ENT, Horodecki-QE}. The resource capabilities of identical particle entanglement were discussed in various works \cite{Killoran2014RCAPENT, Rosario2018IndR, Castellini2019Indist, Gigena2020OneBody, Morris2020RCAPENT, Gigena2021ManyBody}, and some of these works \cite {Gigena2020OneBody, Gigena2021ManyBody} were grounded on Nielsen's theorem. However, as was carefully scrutinized in Ref.~\cite{BENATTI20201}, mode entanglement provides a better framework than particle entanglement for a convenient definition of entanglement in quantum systems composed of indistinguishable particles.

More importantly, indistinguishable systems are subject to superselection rules (SSRs) \cite{PhysRev.88.101}, which impose additional restrictions on the free operations that cannot create any mode entanglement. These constraints, leading to new applications in quantum protocols \cite{2003_PRL_SSRandLOCC, 2004_PRA_SSRandLOCC}, are related to the conservation of physical quantities (Q) such as parity (P) and particle number (N). In particular, Q-SSRs shrink the physical state space of a fermionic mode system by forbidding the states that violate the conservation of Q. This affects the amount of mode entanglement extractable from a given quantum state, which is well understood for bipartite settings \cite{FixedN_2003_PRA, 2013_PRA_fPTrace, 2021_JCTC_DMRG_SSR, Vidal2021-FermionicOps, OrbitalDiscord}. However, to our knowledge, its effect on single-copy state transformations by free operations has been characterized only for the pure entangled states so far \cite{2004_PRL_SSRandLOCC, Schuch2004SSREnt, Barlett2007REVSSR, Gour_2008SSR}. In the first part of this paper (Sec.~\ref{Sec::Conditions}), following Refs.~\cite{2004_PRL_SSRandLOCC, Schuch2004SSREnt}, we provide a set of majorization-based conditions for mixed state transformations of bipartite mode entanglement by LOCCs that are restricted by local SSRs (Algorithm \ref{alg:algoritma}).

Further, Nielsen's theorem allows an ancillary system to play the role of a \emph{catalyst} in entanglement transformations of distinguishable systems \cite{Jonathan1999Catalyst}. Catalyst, which is one of the most pivotal ingredients for QRTs \cite{Anshu2018GRTCatalysts}, enhances our ability to transform one quantum state \(\rho\) into another state \(\sigma\) via free operations \({\mathbb{O}}\). More precisely, with the help of a quantum state \(\tau\), the transformation from \(\rho\) to \(\sigma\) can be replaced by the transformation from \(\rho \otimes \tau\) to \(\sigma \otimes \tau\), where the catalyst \(\tau\) keeps its form at the end of the transformation:
\begin{eqnarray}\label{RhoToSigmaCatalytic}
\rho\otimes\tau \overset{\mathbb{O}}{\longrightarrow} \sigma\otimes\tau.
\end{eqnarray}
The essential point here is that while \(\rho\) cannot be transformed into \(\sigma\) under free operations \({\mathbb{O}}\), the transformation becomes possible when considered as given in Eq.~\eqref{RhoToSigmaCatalytic}, for which \(\tau\) participated as a catalyst. The first attempt of this kind was put forward in Ref.~\cite{Jonathan1999Catalyst}, where it was clearly shown that the catalytic effect can increase the efficiency of entanglement concentration procedures for finite quantum states. It is now well established from a variety of studies \cite{Eisert2000CatMixed, Sumit2001Catalytic, Aberg2014Catalytic, Ng_2015, Kaifeng2016CatCoh, Boes2018Catalytic, Soorya2020CATMAJ, Guo2021, KondraCATALY21, Hyung2021Catalytic, Bartosik2021Catalytic, RubboliCATLY22} that the existence of catalyst allows us to gather more information on the manipulation of resource states.

Our research also aims at extending catalytic transformations to include fermionic mode systems. As an example from quantum chemistry, quantum correlations shared between orbitals can become partially inaccessible due to the SSRs \cite{2021_JCTC_DMRG_SSR, OrbitalDiscord}, which raises doubts about the usefulness of molecular systems as a resource in quantum technologies. Would it be possible to unlock such correlations using ancillary fermionic mode systems as catalysts? The second part of this paper (Sec.~\ref{Sec::Catalyt}) paves the way for the investigation of such question(s).


\section{State Manipulations: Framework} \label{Sec::LOCCPreliminaries}

In this section, to the best of our knowledge, we present a brief overview of our contemporary understanding regarding the manipulation of bipartite entanglement by LOCCs and their interplay with superselection rules. Through this undertaking, we strive to elucidate the unique and novel outcomes of our study, enabling readers to gain a clearer sense of our research findings.

Our ability to control entanglement arises from the limitations imposed by the operations feasible on entangled quantum systems. In this context, LOCC stands out as the inherent class of operations for numerous critical quantum information tasks \cite{Chitambar2014LOCC}. Specifically, for a bipartite quantum system described by Hilbert spaces \( \mathcal{H}_A \) and \( \mathcal{H}_B \) associated with subsystems \(A\) and \(B\), respectively, an operation is deemed LOCC if it can be expressed as a composition of local operations acting independently on each subsystem, followed by classical communication. Generally speaking, the fundamental question in the resource theory of bipartite entanglement revolves around characterizing the set of states that can be converted into one another under LOCC and understanding the entanglement properties during such transformations.

Majorization has been proven to be an effective means at characterizing and manipulating bipartite entanglement \cite{Bhatia, Marshall}. 
Formally, given two probability distributions \(\mathbf{p}\) and \(\mathbf{q}\) defined on a finite set of probabilities, \(\mathbf{p}\) is said to be majorized by \(\mathbf{q}\) (written \(\mathbf{p} \prec \mathbf{q}\)) if and only if the cumulative sums of the probabilities in \(\mathbf{q}\) are greater than or equal to the cumulative sums of the probabilities in \(\mathbf{p}\). This can be represented as
\begin{equation}
\sum_{i=1}^{k} p_{i}^\downarrow \leq \sum_{i=1}^{k} q_{i}^\downarrow,
\end{equation}
for all \(k = 1, 2, \ldots, n\), where \(\{p_{i}^\downarrow\}_i\) and \(\{q_{i}^\downarrow\}_i\) denote the probabilities arranged in non-increasing order. Regarding the matter of utmost interest to us, it was demonstrated that a bipartite pure quantum state \(\rho\) is said to be majorized by another pure state \(\sigma\), written \(\rho \prec \sigma\), if and only if the eigenvalues of the reduced density matrices of \(\rho\) and \(\sigma\) satisfy the majorization relation \cite{Nielsen-Maj, Vidal2001InterConv}, written 
\begin{equation}
\lambda(\rho_A) \prec \lambda(\sigma_A),
\end{equation}
where \(\lambda(\rho_A)\) and \(\lambda(\sigma_A)\) are the eigenvalue sequences of the reduced density matrices of \(\rho\) and \(\sigma\) on subsystem \(A\), respectively. Importantly, the connection between majorization theory and bipartite entanglement stems from the observation that an increase in entanglement within a pure system necessitates a greater degree of disorder in its subsystems \cite{TJP2023}. Therefore, if a bipartite pure state \(\rho\) can be transformed into another pure state \(\sigma\) through LOCC transformations, then \(\rho \prec \sigma\) holds. This implies that the majorization relation between the eigenvalue sequences of the initial and final states is preserved under LOCC operations. Precisely, for any LOCC transformation, we have \cite{Nielsen-Maj, Vidal2001InterConv}:
\begin{equation}
\rho \overset{\text{LOCC}}{\longrightarrow} \sigma \quad \iff \quad \rho \prec \sigma.
\end{equation}

On the other hand, SSRs establish constraints on the specific properties within physical systems \cite{Barlett2007REVSSR}. Namely, these rules are particularly significant in the context of particle number and parity, influencing the dynamics of entanglement and its manipulation~\cite{Bartlett2003EntConstSSRs}. When considering systems subject to SSRs, such as those with fixed particle number or definite parity, certain entanglement transformations may become inaccessible or restricted. These limitations arise due to the inherent restrictions on free operations that can be performed within the context of SSRs, thereby shaping the landscape of entanglement-free manipulation and entanglement-based quantum protocols. This issue has been subject to meticulous examination in various studies \cite{Bartlett2006EntUnderRestOp, Jones2006EntAndSymmetry, MariCarmen2009EntFermions, White2009AccessEntSSRs, Hatem2018AccessEntEntropy}. For instance, Schuch \emph{et al}. \cite{2004_PRL_SSRandLOCC, Schuch2004SSREnt} proposed some measures to quantify the nonlocal properties of quantum entanglement in the presence of SSRs, focusing on both the formation and distillation aspects. These measures establish an order for quantum states, ensuring that the asymptotic conversion of two states through LOCC is achievable only when the measures decrease throughout the transformation. Additionally, the study in \cite{2004_PRL_SSRandLOCC, Schuch2004SSREnt} explored the role of majorization pre-order in the single copy transformation of pure entangled states. In the next section, we will revisit these results and compare them with our own findings.

In our endeavor to uncover fresh insights beyond the existing achievements reported in the literature, we investigate the single copy transformation of mixed entangled states in the presence of local parity SSR. By employing an ancillary mode system that remains unchanged in the transformation, we investigate the feasibility of inducing a catalytic effect of mode entanglement on the local parity. Our study presents compelling evidence for the catalytic capacity of the ancillary mode system, offering a pathway to circumvent the restrictions imposed by the local parity SSR. From this perspective, we believe that our study contributes new results that enhance the existing literature. In the subsequent section, we elaborate on these results.


\section{Conditions for Fermionic Mode Entanglement Transformations} \label{Sec::Conditions}

\begin{figure}[t]
	\centering
	\includegraphics[width=.92\columnwidth]{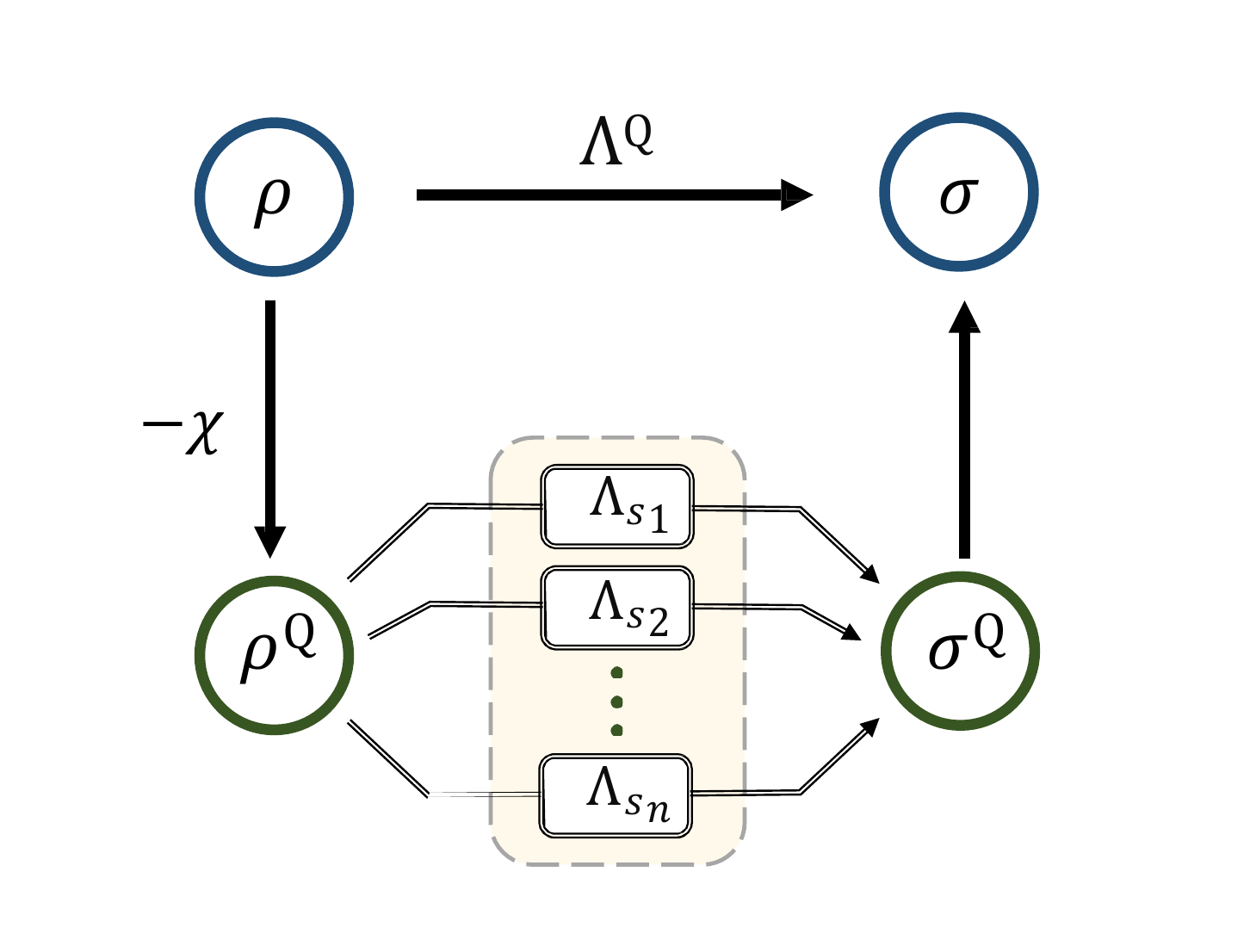}
	\caption{Fermionic mode state transformation from \(\rho\) to \(\sigma\) under LOCC that is restricted by local Q-SSR, with Q being P for parity and N for particle number. Here, the downward thick arrow illustrates the separation of sectors \(s_j\) of Fock space due to local Q-SSR, while excluding the term \(\chi\). This term pertains to the portion of superposition that is not useful for free operations and should be eliminated during the transformation. The subsequent horizontal (thin) arrows depict LOCC transformations within sectors that are independent of each other, such that \(\rho_{s_j} \mapsto \Lambda_{s_j}(\rho_{s_j})=\sigma_{s_j}\). This stage involves the manipulation of the resource. Finally, the upward thick arrow signifies the equality \(\sigma = \sigma^Q\). A detailed analysis of our consideration is provided in the main text.}
	\label{Fig:FerCatTrs}
\end{figure}

Throughout this section, we consider a fermionic system that consists of two-mode subsystems called \emph{orbital}s. We also assume that each mode inside the same orbital corresponds to a different \emph{spin} value. Furthermore, total parity, total particle number, and total spin in the whole system are taken to be constant.

A given local SSR splits the fermionic Fock space into different sectors, labeled by \(s_j\), such that any superposition of the states that live in different sectors is not allowed for a physical system. For the sake of simplicity and without loss of generality, let us focus on the case where the total parity is even, the total number of electrons is two, and the total spin is zero. Then, the fermionic Fock space is split into two distinct sectors under local P-SSR. Namely, one sector has orbitals where local parity is even, that is, \(\{\ket{00,11}, \ket{11,00}\}\), denoted by \(s_1=s_{ee}\), and the other sector has orbitals where local parity is odd, that is, \(\{\ket{01,10}, \ket{10,01}\}\), denoted by \(s_2=s_{oo}\). In this case, we can write an initial resource state \(\rho\) such that
 \begin{eqnarray}\label{RHO-S1andS2} \rho \equiv \rho(P, p_1, p_2, \alpha_1, \alpha_2) = P \rho_{s_1} + \big(1 - P\big)\rho_{s_2} + \chi_{\rho},
\end{eqnarray}
where the inter-sector probabilities are represented by uppercase letters, i.e., \(P_{s_1} = P\) and \(P_{s_2} = 1 - P\), lowercase letters \(p_1\) and \(p_2\) stand for the intra-sector probabilities, and Greek letters \(\alpha_1\) and \(\alpha_2\) are responsible for the intra-sector quantum coherence terms. Then the projections of \(\rho\) given in Eq.~\eqref{RHO-S1andS2} onto the sectors \(s_1\) and \(s_2\) are respectively given by
\begin{eqnarray}\label{RHO-S1}
\rho_{s_1}
&=& p_1 \ket{00,11}\bra{00,11} + \big(1-p_1\big) \ket{11,00}\bra{11,00} \nonumber \\
& & + \   \alpha_1 \ket{00,11}\bra{11,00}
+ \alpha^*_1 \ket{11,00}\bra{00,11}
\end{eqnarray}
and
\begin{eqnarray}\label{RHO-S2}
\rho_{s_2} &=& p_2 \ket{01,10}\bra{01,10} + \big(1-p_2\big) \ket{10,01}\bra{10,01} \nonumber \\
& & + \   \alpha_2 \ket{01,10}\bra{10,01}+\alpha^*_2 \ket{10,01}\bra{01,10}.
\end{eqnarray}

In Eq.~(\ref{RHO-S1andS2}), \(\chi_{\rho}\) comprises superposition terms that can be exclusively utilized through LOCC in the absence of P-SSR. Unfortunately, when P-SSR is present, our ability to control this resource diminishes, rendering it inaccessible for use and effectively lost. The matrix elements of \(\chi_{\rho}\) correspond to cross terms of \(\{\ket{00,11}, \ket{11,00}\}\) and \(\{\ket{01,10}, \ket{10,01}\}\). To construct the \textit{accessible}  part of the state under local P-SSR constraints, we need to exclude these terms and denote the remaining terms in Eq.~\eqref{RHO-S1andS2} as \(\rho^{\text{P}} = \rho - \chi_\rho\), as shown in Fig.~\ref{Fig:FerCatTrs}.

The same decomposition also holds for the target resource state \(\sigma\), which reads
\begin{eqnarray}\label{SIGMA-S1andS2}
\sigma \equiv \sigma(Q, q_1, q_2, \beta_1, \beta_2) = Q \sigma_{s_1} + (1 - Q)\sigma_{s_2} + \chi_{\sigma},
\end{eqnarray}
where its projections onto the sectors \(\sigma_{s_1}\) and \(\sigma_{s_2}\) are obtained by replacing \(\rho\) with \(\sigma\), \(P\) with \(Q\), \(p_j\) with \(q_j\), and \(\alpha_j\) with \(\beta_j\) (for \(j=1,2\)) in Eqs.~\eqref{RHO-S1}~and~\eqref{RHO-S2}.

First of all, it is worth emphasizing that SSR-restricted LOCC must commute with the operator \(\hat{Q}\) to avoid violating local Q-SSR. This requirement makes it block-diagonal in the relevant sectors, similar to \(\hat{Q}\). Therefore, the transformation \(\rho \mapsto \Lambda^{\text{P}}(\rho)=\sigma\) should be described by \(\hat{K} \, \rho \, \hat{K}^\dagger\), where \(\hat{K} = \hat{K}_{s_1} + \hat{K}_{s_2}\). Given that \(\hat{K}_{s_j} \, \rho_{s_{k \neq j}} \, \hat{K}_{s_j}^\dagger = \hat{K}_{s_j} \, \chi_\rho \, \hat{K}_{s_j}^\dagger = 0\), it is straightforward to demonstrate that 
\begin{equation}\label{Eq::OpenTransform}
    \rho \mapsto \Lambda^{\text{P}}(\rho) = P \, \hat{K}_{s_1} \, \rho_{s_1} \hat{K}_{s_1}^\dagger + \big(1 - P\big) \hat{K}_{s_2} \, \rho_{s_2} \hat{K}_{s_1}^\dagger .
\end{equation}

Eq.~\eqref{Eq::OpenTransform} indicates that P-SSR compatible LOCC operations must have a separate effect on each sector of the fermionic Fock space. More precisely, free operations should not be able to transfer any information between the sectors \(s_1\) and \(s_2\) during the transformation. Then, for all sectors \(s_j\), the equality \(P_{s_j} = Q_{s_j}\) must hold, and a LOCC denoted by \(\Lambda_{s_j}(\rho_{s_j}) = \hat{K}_{s_j} \, \rho_{s_j} \, \hat{K}_{s_j}\) must exist in order to achieve the transformation \(\rho_{s_j} \mapsto \sigma_{s_j}\) (see Fig.~\ref{Fig:FerCatTrs}). These two conditions have been also considered in \cite{2004_PRL_SSRandLOCC, Schuch2004SSREnt} for the single copy transformation of pure entangled states. However, they are necessary but not sufficient. In addition to them, the superposition terms \(\chi\) should vanish during the transformation as demonstrated in \eqref{Eq::OpenTransform}. With \(\sigma-\sigma^{\text{Q}}=\chi_{\sigma}\), the equality \(\chi_{\sigma} = 0\) must be satisfied. In other words, the target state \(\sigma\) cannot be pure unless it exists within a single sector. This implies that contrary to the claim in Refs.~\cite{2004_PRL_SSRandLOCC, Schuch2004SSREnt}, single-copy transformations of generic pure-states cannot be verified using majorization. Then the transformation \(\Lambda^{\text{Q}}(\rho)\) in Eq.~\eqref{Eq::OpenTransform} can be mathematically rewrtitten as
\begin{eqnarray}
\left\{\Lambda(\rho^{\text{Q}})=\sum_{s_j} P_{s_j} \Lambda_{s_j}(\rho_{s_j}) \ ; \ \Lambda(\chi_{\rho}) = 0\right\}.
\end{eqnarray}

To check the presence of \(\Lambda_{s_j}\) that denotes an intra-sector LOCC, we must first take the fermionic partial trace~\cite{2013_PRA_fPTrace} of \(\rho_{s_j}\) and \(\sigma_{s_j}\) over the modes of one orbital (subsystem). The eigenvalues of the resulting reduced density matrices, namely the \textit{Schmidt coefficients}, are required to construct the following probability vectors:
\begin{eqnarray}\label{sectors-Rho}
\mathbb{S}_{s_1}(\rho_B^{\text{P}})= \big\{p_1, 1-p_1\big\}\,  ; \quad
\mathbb{S}_{s_2}(\rho_B^{\text{P}})= \big\{p_2, 1-p_2\big\},
\end{eqnarray}
\begin{eqnarray}\label{sectors-Sigma}
\mathbb{S}_{s_1}(\sigma_B^{\text{P}})= \big\{q_1, 1-q_1\big\}\,  ; \quad
\mathbb{S}_{s_2}(\sigma_B^{\text{P}})= \big\{q_2, 1-q_2\big\}.
\end{eqnarray}
Then, Nielsen's theorem \cite{Nielsen-Maj} can be applied by using these marginal probability vectors. The presence of \(\Lambda_{s_1}\) and \(\Lambda_{s_2}\) are verified if the majorization conditions \(\mathbb{S}_{s_1}(\rho_B^{\text{P}}) \prec \mathbb{S}_{s_1}(\sigma_B^{\text{P}})\) and \(\mathbb{S}_{s_2}(\rho_B^{\text{P}}) \prec \mathbb{S}_{s_2}(\sigma_B^{\text{P}})\) are both satisfied.

Embodying this example, we can propose a set of conditions for bipartite mode entanglement transformation \(\rho \mapsto \Lambda^{\text{P}}(\rho)=\sigma\) as in Algorithm~\ref{alg:algoritma} in Supplementary Material. Initial and target resource states \(\rho\) and \(\sigma\) do not have to be pure for this algorithm to work correctly. However, the last step of the algorithm requires that the projections of these states onto the sectors, namely \(\rho_{s_j}\) and \(\sigma_{s_j}\) must be pure, i.e., \(|\alpha_j|^2 = p_j (1-p_j)\) and \(|\beta_j|^2 = q_j (1-q_j)\). Otherwise, it would not be possible to analyze the transformations \(\rho_{s_j} \mapsto \Lambda_{s_j}(\rho_{s_j})=\sigma_{s_j}\) with respect to the majorization criteria. If the target state of the fermionic system is not constrained to remain within a single sector, it is not possible for both \(\sigma\) and \(\sigma_{s_j}\) to be pure simultaneously as this state cannot include any inter-sector superposition \(\chi_\sigma\). Therefore, we cannot use majorization in this manner for the LOCC transformation of generic pure states. Unfortunately, this has not been recognized in \cite{2004_PRL_SSRandLOCC, Schuch2004SSREnt}.

To conclude, the generalization of majorization in the case of fermionic systems that we proposed here for local P-SSR is also valid for local N-SSR. This time, the fermionic Fock space belonging to the two-electron two-orbital systems above split into three sectors instead of two: That is, zero-two local particle-number sector, one-one local particle-number sector, and two-zero local particle-number sector.

\section{Catalytic Transformations of Bipartite Mode Entanglement}\label{Sec::Catalyt}

\subsection{Catalytic transformations}\label{SubSec::Cat-Transform}

There are three scenarios in which a local SSR-restricted LOCC transformation may not be possible, as delineated by the set of conditions specified in Algorithm ~\ref{alg:algoritma}. The possible algorithmic failures at lines 28, 25, and 22 correspond to the first, second, and third scenarios, respectively. However, the inclusion of a catalyst as in Eq.~\eqref{RhoToSigmaCatalytic} may drive an otherwise impossible fermionic state transformation. We can exemplify this possibility through the four-mode system described in Sec.~\ref{Sec::Conditions}.

Let us first discuss the appropriate dimension of \(\tau\) involved in the transformation. The catalyst \(\tau\) cannot have two modes, since the local P-SSR prohibits any quantum coherence in its state in this case. Also, a catalyst with more than four modes would be trivial as it becomes bigger than the original system. Hence, it is reasonable to choose the catalytic state to be
\begin{eqnarray}\label{catalyst} 
\tau = R \, \tau_{e} + (1-R) \tau_{o} +\chi_{\tau}, 
\end{eqnarray}
where \(\tau_{e}\) (\(\tau_{o}\)) shares the same form with \(\rho_{e}\) and \(\sigma_{e}\) (\(\rho_{o}\) and \(\sigma_{o}\)) but has different parameters for the probability and coherence inside the even-even (odd-odd) sector of the Fock space, e.g., \(r_1\) (\(r_2\)) and \(\gamma_1\) (\(\gamma_2\)).

{\bf{Example 1}}: \emph{A bipartite mode entanglement transformation that cannot be catalyzed by \(\tau\)}.

In the current scenario, let us consider that the initial two conditions given in Algorithm \ref{alg:algoritma} are fulfilled, while the final condition remains unfulfilled: that is, \(\chi_{\sigma} = 0\), \(P = Q = S\), \(\mathbb{S}_{s_1}(\rho_B^{\text{P}}) \prec \mathbb{S}_{s_1}(\sigma_B^{\text{P}})\), and \(\mathbb{S}_{s_2}(\rho_B^{\text{P}}) \succ \mathbb{S}_{s_2}(\sigma_B^{\text{P}})\). After adding the catalyst, we must run Algorithm~\ref{alg:algoritma} for the initial and target states of the joint system, that are \(\rho^{\prime} = \rho\wedge\tau\) and \(\sigma^{\prime} = \sigma\wedge\tau\), to check if the conversion from \(\rho^{\prime}\) into \(\sigma^{\prime}\) is possible. Note that it is not the tensor product (\(\otimes\)) but the wedge product (\(\wedge\)) that forms a larger fermionic Fock space for the joint system from the Fock spaces of the two-orbital system and the catalyst.

The first condition is automatically fulfilled in the case of \(\chi_{\tau}=0\). Also, it is straightforward to show that the second condition is still satisfied in this case. As the total parity does not change after the addition of the catalyst, the joint Fock space is divided into two sectors too. However, there are more than two vectors spanning each sector this time. Nevertheless, the inter-sector probabilities in the initial and target states remain the same, that is, \(P_{s_2}^\prime = Q_{s_2}^\prime = S + R - 2\, S R\) and \(P_{s_1}^\prime = Q_{s_1}^\prime = 1 - P_{s_2}^\prime\).

On the other hand, we have yet to verify the third condition, which requires the pure projections of the initial and target quantum states in order to facilitate a comparison based on the majorization condition. Projections of \(\rho^\prime\) and \(\sigma^\prime\) onto the even-even and odd-odd sectors are not pure unless \(R\) is either 0 or 1. However, neither of these choices allow us to obtain the same majorization preorder inside both sectors. To instantiate this, let us suppose \(R = 1\). In this case, the probability vector \(\mathbb{S}_{s_j}({\rho^\prime}_{B^\prime}^{\text{P}})\) turns out to be
\begin{equation}\label{RhoSj4NoCatalA}
\big\{p_j r_1, p_j (1-r_1),(1-p_j) \, r_1, (1-p_j)(1-r_1)\big\},
\end{equation}
whereas \(\mathbb{S}_{s_j}({\sigma^\prime}_{B^\prime}^{\text{P}})\) becomes
\begin{eqnarray}\label{SigmaSj4NoCatalA}
\big\{q_j r_1, q_j (1-r_1),(1-q_j) \, r_1, (1-q_j)(1-r_1)\big\}.
\end{eqnarray}
When compared to Eqs.~\eqref{sectors-Rho}~and~\eqref{sectors-Sigma}, the probability vectors above imply that \(\tau\) given in Eq.~\eqref{catalyst} is not sufficient to change the majorization preorder inside only a single sector. We conclude our discussion on the first example at this juncture and proceed with the second example.

{\bf{Example 2}}: \emph{A bipartite mode entanglement transformation that can be catalyzed by \(\tau\)}.

We start by assuming that the uncatalyzed transformation is not likely to occur as it requires the change of inter-sector probabilities from \(P_{s_j}\) to \(Q_{s_j}\), though the first and last conditions in Algorithm~\ref{alg:algoritma} are satisfied. Once again, we set \(\chi_{\tau}=0\) in order to verify the first condition which implies \(\chi_{\sigma^\prime}=0\). After that, the inter-sector probabilities read \(1 - R - S \, (1 - 2 R)\) and \(S + R - 2 S R\), respectively for the projections onto the even-even and odd-odd sectors. Here, \(S=P\) (\(S=Q\)) for the initial (target) joint state. It is possible to satisfy the second condition if \(R = 1/2\), which in turn gives a probability of \(1/2\) for each projection. 
Remember that the last condition in Algorithm~\ref{alg:algoritma} requires pure projections onto the even-even and odd-odd sectors, i.e., \(\text{tr}[(\rho^\prime_{s_j})^2]\) and \(\text{tr}[(\sigma^\prime_{s_j})^2]\) should equal unity. This can be ensured consistently with the initial assumption of \(P_{s_j} \neq Q_{s_j}\) only when \(P = 0\) and \(Q = 1\) or \textit{vice versa} as the trace of the square of all projections equals \(1 - 2 S \, (1 - S)\), where \(S\) is \(P\) (\(Q\)) for \(\rho^\prime_{s_j}\) (\(\sigma^\prime_{s_j}\)).

Let us proceed with the assumption of \(P = 1\) and \(Q = 0\). In this case, the probability vector \(\mathbb{S}_{s_j}({\rho^\prime}_{B^\prime}^{\text{P}})\) becomes
\begin{equation}\label{RhoSj4NoCatalB}
\left\{p_1 r_j, p_1 (1-r_j),(1-p_1) \, r_j, (1-p_1)(1-r_j)\right\},
\end{equation}
and it can be majorized by \(\mathbb{S}_{s_j}({\sigma^\prime}_{B^\prime}^{\text{P}})\), which reads
\begin{eqnarray}\label{SigmaSj4NoCatalB}
\left\{q_2 r_j, q_2 (1-r_j),(1-q_2) \, r_j, (1-q_2)(1-r_j)\right\}.
\end{eqnarray}
Hence, with the help of a catalyst state, it is possible to change the local parity under Q-SSR-restricted LOCC. This remarkable catalysis can be exemplified, for instance, by the conversion of the given initial resource state
\begin{eqnarray}\label{example-RHO}
\rho = \ket{\psi} \bra{\psi} \  \  \text{s.t.} \   \  \ket{\psi} = 0.4 \ket{00,11} + \sqrt{0.84} \ket{11,00},
\end{eqnarray}
to the target resource state
\begin{eqnarray}\label{example-SIGMA}
\sigma = \ket{\phi} \bra{\phi} \   \  \text{s.t.} \  \  \ket{\phi} = 0.3 \ket{01,10} + \sqrt{0.91} \ket{10,01},
\end{eqnarray}
in the presence of a catalyst that exists in the following state
\begin{eqnarray}\label{example-TAU}
\tau = \frac{1}{2}\left(\ket{\Psi}\bra{\Psi} + \ket{\Phi}\bra{\Phi}\right).
\end{eqnarray}
Here, the states \(\ket{\Psi}\) and \(\ket{\Phi}\) seen in Eq.~\eqref{example-TAU} are given such that
\begin{eqnarray}\label{example-TAU-Psi}
\ket{\Psi} = 0.5 \ket{00,11} + \sqrt{0.75} \ket{11,00}
\end{eqnarray}
and 
\begin{eqnarray}\label{example-TAU-Phi}
\ket{\Phi} = 0.5 \ket{01,10} + \sqrt{0.75} \ket{10,01}.
\end{eqnarray}
For this explicit example, it is straightforward to show that \(\mathbb{S}_{s_j}\big(\text{tr}_{B^\prime}[\sigma^\prime]\big) \succ \mathbb{S}_{s_j}\big(\text{tr}_{B^\prime}[\rho^\prime]\big)\) for both sectors, where 
\begin{eqnarray}\label{FinalSectors-initial}
\mathbb{S}_{s_j}\big(\text{tr}_{B^\prime}[\rho^\prime]\big) = \big\{0.04, 0.12, 0.21, 0.63\big\}
\end{eqnarray}
and 
\begin{eqnarray}\label{FinalSectors-target}
\mathbb{S}_{s_j}\big(\text{tr}_{B^\prime}[\sigma^\prime]\big) = \big\{0.0225, 0.0675, 0.2275, 0.6825\big\},
\end{eqnarray}
for \(j=1, 2\) (i.e., both sectors comprise identical distributions). As a consequence, \(\mathbb{S}_{s_j}\big(\text{tr}_{B^\prime}[\sigma^\prime]\big)\) given in Eq.~\eqref{FinalSectors-target} majorizes \(\mathbb{S}_{s_j}\big(\text{tr}_{B^\prime}[\rho^\prime]\big)\) given in Eq.~\eqref{FinalSectors-initial}. Therefore, an impossible bipartite mode entanglement transformation under local SSRs and LOCC can be catalyzed by another fermionic system consisting of the same number of modes.

In the following discussion, we initially focus on presenting the general structure. Once we have established the general foundation in Sec.~\ref{SubSec::Fermionic-Operations} and Sec.~\ref{SubSec::Explicit-Construct}, we revisit this specific example for further analysis in Sec.~\ref{SubSec::Example2}.

\subsection{Local fermionic quantum operations}\label{SubSec::Fermionic-Operations}

We now provide the explicit forms of the local fermionic quantum operations \cite{Vidal2021-FermionicOps} that enable the transformation from \(\rho'\) into \(\sigma'\), expressed in the form of 
\begin{eqnarray}\label{Transf-ExplicitQOper}
\rho' = \rho \wedge \tau \overset{\text{Q-SSR}}{\longrightarrow} \sigma' = \sigma \wedge \tau.
\end{eqnarray}
By presenting local operations, we aim to enhance transparency and facilitate a more straightforward understanding of the underlying (fermionic) quantum processes involved in this transformation given by Eq.~\eqref{Transf-ExplicitQOper}.

For brevity, we introduce the following notation
\begin{eqnarray}\label{e-12345678}\begin{aligned}
& \ket{e_{1}} := \ket{0000,1111}, \quad 
\ket{e_{2}} := \ket{0011,1100}, &  \\
& \ket{e_{3}} := \ket{1100,0011}, \quad
\ket{e_{4}} := \ket{1111,0000}, &  \\
& \ket{e_{5}} := \ket{0101,1010}, \quad 
\ket{e_{6}} := \ket{0110,1001}, &  \\
& \ket{e_{7}} := \ket{1001,0110}, \quad
\ket{e_{8}} := \ket{1010,0101}, & 
\end{aligned}\end{eqnarray}
and
\begin{eqnarray}\label{o-12345678}\begin{aligned}
& \ket{o_{1}} := \ket{0001,1110}, \quad 
\ket{o_{2}} := \ket{0010,1101}, &  \\
& \ket{o_{3}} := \ket{1101,0010}, \quad
\ket{o_{4}} := \ket{1110,0001}, &  \\
& \ket{o_{5}} := \ket{0100,1011}, \quad 
\ket{o_{6}} := \ket{0111,1000}, &  \\
& \ket{o_{7}} := \ket{1000,0111}, \quad
\ket{o_{8}} := \ket{1011,0100}. & 
\end{aligned}\end{eqnarray}
This notation, specifically denoted as \(\{\ket{e_{i}}\}_{i=1}^{8}\) for even parity sector and \(\{\ket{o_{i}}\}_{i=1}^{8}\) for odd parity sector, makes it much easier to follow our discussion. Expressed as
\(\ket{e_{i}} \equiv \ket{e_{A',i}} \wedge \ket{e_{B',i}}\) and  \(\ket{o_{i}} \equiv \ket{o_{A',i}} \wedge \ket{o_{B',i}}\), these further serve as a compact form for Eqs.~\eqref{e-12345678} and \eqref{o-12345678}. In this setting, \(\ket{e_{A',2}}\) (and \(\ket{e_{B',3}}\)) is, for instance, equal to \(\ket{0011}\)

The transformation of interest can be realized with a unit probability by employing two distinct sets of local fermionic operators. Thus, the transformation \eqref{Transf-ExplicitQOper} can be analyzed as comprising two sequential steps. The first step is characterized by the utilization of local operators
\begin{eqnarray}\label{Local-Op-Epsilon-A}
\hat{\mathcal{E}}_{A'}^{k} &=& \sqrt{\mathfrak{p}_k} \sum_{i=1}^{4} f_{k}(i) \; \pi_{i}^{k} \ket{e_{A',i}}\bra{e_{A',i}} \nonumber \\ & & \oplus 
\sqrt{\mathfrak{p}_k} \sum_{j=1}^{4}\ket{o_{A',j}}\bra{o_{A',j}}
\end{eqnarray}
and
\begin{eqnarray}\label{Local-Op-Epsilon-B}
\hat{\mathcal{E}}_{B'}^{k} = \sum_{i=1}^{4} \; \pi_{i}^{k} \ket{e_{B',i}}\bra{e_{B',i}} \oplus 
\sum_{j=1}^{4}\ket{o_{B',j}}\bra{o_{B',j}}.
\end{eqnarray}
Here, \(\{\mathfrak{p}_k\}\) are the probabilities of the outcomes, subject to the condition \(\sum_{k=1}^{4} \mathfrak{p}_k = 1\); the function \(f_{k}(i)\) is defined based on the coefficients of \(\rho'\) and \(\sigma'\); and \(\{\pi_{i}^{k}\}\) represents permutations, where the precise determination thereof is crucial for overall analysis. In addition, the local operators \(\{\hat{\mathcal{E}}_{B'}^{k}\}\) given by Eq.~\eqref{Local-Op-Epsilon-B} are defined as unitary transformations acting on subsystem \(B'\) \cite{Vidal2021-FermionicOps}. The operation with the local operators \(\{\hat{\mathcal{E}}_{A'}^{k}\}\) and \(\{\hat{\mathcal{E}}_{B'}^{k}\}\) operates specifically on the even sector of the initial state \(\rho'\), leading to a transformation that aligns it with the corresponding even sector of the target state \(\sigma'\). Concurrently, and importantly, the odd sector of the initial state remains unaffected. Therefore, the transformation in the first step is expressed as:
\begin{eqnarray}\label{First-Step-varRho2}
\mathbb{E}({e_{\rho'} \rightarrow e_{\sigma'}}) : \;  \sum_{k=1}^{4} \Big[\hat{\mathcal{E}}_{A'}^{k} \wedge \hat{\mathcal{E}}_{B'}^{k}\Big] \rho' \Big[\hat{\mathcal{E}}_{A'}^{k} \wedge \hat{\mathcal{E}}_{B'}^{k}\Big]^{\dagger} = \varrho.
\end{eqnarray}
Thus, while the elements comprising the even sector of the intermediate state \(\varrho\) in Eq.~\eqref{First-Step-varRho2} correspond to the even sector elements of the target state \(\sigma'\), the elements shaping the odd sector align with the odd sector elements of the initial state \(\rho'\). 

Following the completion of the transformation in the first step, the second step is characterized by the utilization of local operators given by
\begin{eqnarray}\label{Local-Op-O-A}
\hat{\mathcal{O}}_{A'}^{k} &=&
\sqrt{\mathfrak{q}_k}\sum_{i=5}^{8}\ket{e_{A',i}}\bra{e_{A',i}} \nonumber \\ & &  \oplus \sqrt{\mathfrak{q}_k} \sum_{j=1}^{4} g_{k}(j) \; \pi_{j}^{k} \ket{o_{A',j}}\bra{o_{A',j}}
\end{eqnarray}
and
\begin{eqnarray}\label{Local-Op-O-B}
\hat{\mathcal{O}}_{B'}^{k} = \sum_{i=5}^{8} \; \ket{e_{B',i}}\bra{e_{B',i}} \oplus 
\sum_{j=1}^{4}\pi_{j}^{k}\, \ket{o_{B',j}}\bra{o_{B',j}}.
\end{eqnarray}
Here, \(\{\mathfrak{q}_k\}\) are the probabilities of the outcomes, subject to the condition \(\sum_{k=1}^{4} \mathfrak{q}_k = 1\); the function \(g_{k}(j)\) is defined based on the coefficients of the intermediate state \(\varrho\) and target state \(\sigma'\); and \(\{\pi_{j}^{k}\}\) represents permutations as usual. In addition, the local operators \(\{\hat{\mathcal{O}}_{B'}^{k}\}\) given by Eq.~\eqref{Local-Op-O-B} are defined as unitary transformations acting on subsystem \(B'\). The operation with the local operators \(\{\hat{\mathcal{O}}_{A'}^{k}\}\) and \(\{\hat{\mathcal{O}}_{B'}^{k}\}\) operates specifically on the odd sector of the intermediate state \(\varrho\), leading to a transformation that aligns it with the corresponding odd sector elements of the target state \(\sigma'\). While this occurs, the even sector of \(\varrho\) remains unaffected. Thus, the transformation in the second step is expressed as:
\begin{eqnarray}\label{Second-Step-varRho2}
\mathbb{O}({o_{\rho'} \rightarrow o_{\sigma'}}) : \,  \sum_{k=1}^{4} \Big[\hat{\mathcal{O}}_{A'}^{k} \wedge \hat{\mathcal{O}}_{B'}^{k}\Big] \varrho \Big[\hat{\mathcal{O}}_{A'}^{k} \wedge \hat{\mathcal{O}}_{B'}^{k}\Big]^{\dagger} = \sigma'.
\end{eqnarray}
Consequently, the transition from the initial state \(\rho'\) to the target state \(\sigma'\) is accomplished by performing the fermionic local quantum operations presented in Eqs.~\eqref{First-Step-varRho2} and \eqref{Second-Step-varRho2} successively as explained.

The remaining task pertains exclusively to the determination of probabilities. To obtain the probabilities \(\{\mathfrak{p}_k\}\) and \(\{\mathfrak{q}_k\}\), we need to solve the equations
\begin{eqnarray}
\sum_{k=1}^{4} \big(\hat{\mathcal{E}}_{A'}^{k}\big)^{\dagger}  \hat{\mathcal{E}}_{A'}^{k} = \mathds{1}, \quad 
\sum_{k=1}^{4} \big(\hat{\mathcal{O}}_{A'}^{k}\big)^{\dagger}  \hat{\mathcal{O}}_{A'}^{k} = \mathds{1},
\end{eqnarray}
respectively. Importantly, solving our problem depends on accurately figuring out the permutations, and therefore, the functions \(f_k(i)\) and \(g_k(j)\). Once we do that, we can determine the exact probabilities. In that connection, in Supplemental Material, we give an exhaustive presentation of the permutations and 
probabilities, offering a detailed account of their respective arrangements and values.

\subsection{Explanatory insight into catalytic transformation components}\label{SubSec::Explicit-Construct}

In this section, we thoroughly explain the ingredients of transformation \eqref{Transf-ExplicitQOper}, namely, the explicit depiction of the even sector and the odd sector portions. The given initial state \(\rho=\ket{\psi}\bra{\psi}\) is characterized by the normalized state
\begin{equation}
\ket{\psi} = \psi_1 \ket{00,11} + \psi_2 \ket{11,00}, \quad \big(\psi_i \in \mathbb{R}\big).
\end{equation}
Similarly, the target state \(\sigma=\ket{\phi}\bra{\phi}\) is characterized by the normalized state \(\ket{\phi}\), which is described by 
\begin{equation}
\ket{\phi} = \phi_1 \ket{01,10} + \phi_2 \ket{10,01}, \quad \big(\phi_i \in \mathbb{R}\big).
\end{equation}
In addition to these states, we introduce \(\ket{\Psi} = \Psi_1 \ket{00,11} + \Psi_2 \ket{11,00}\) (\(\Psi_i \in \mathbb{R}\)) and \(\ket{\Phi} = \Phi_1 \ket{01,10} + \Phi_2 \ket{10,01}\) (\(\Phi_i \in \mathbb{R}\)) as relevant entities --- serving as the bases of the catalyst --- for our analysis. The expression of the initial state, augmented by the wedge product of \(\rho\) and \(\tau\), is given by
\begin{eqnarray}\label{Rho'RhotimesTau}
\rho' =  \rho \wedge \tau = \frac{1}{2}\big(e_{(\rho')} + o_{(\rho')}\big).    
\end{eqnarray}
Then, following a series of mathematical calculations, we deduce that 
\begin{widetext}
\begin{eqnarray}\label{Explicit-e-rho}
e_{(\rho')} 
&=& {\Psi_1^2} \bigg[\Big(\psi_1 \ket{e_1} + \psi_2 \ket{e_3}\Big)\Big(\psi_1 \bra{e_1} + \psi_2 \bra{e_3}\Big)\bigg] + {\Psi_2^2} \bigg[\Big(\psi_1 \ket{e_2} + \psi_2 \ket{e_4}\Big)\Big(\psi_1 \bra{e_2} + \psi_2 \bra{e_4}\Big)\bigg]  \nonumber \\
&& + {\Psi_1\Psi_2} \bigg[\psi_1\psi_2\Big(\ket{e_1}\bra{e_4} + \ket{e_4}\bra{e_1} + \ket{e_2}\bra{e_3} + \ket{e_3}\bra{e_2}\Big) + \psi_1^2 \Big(\ket{e_1}\bra{e_2} +\ket{e_2}\bra{e_1}\Big) + \psi_2^2 \Big(\ket{e_3}\bra{e_4} +\ket{e_4}\bra{e_3}\Big)\bigg]. \;  \nonumber \\
\end{eqnarray}
\end{widetext}
Further simplification of the expression in Eq.~\eqref{Explicit-e-rho} is possible by introducing \(\ket{\eta_1} = \psi_1 \ket{e_1} + \psi_2 \ket{e_3}\) and
\(\ket{\eta_2} = \psi_1 \ket{e_2} + \psi_2 \ket{e_4}\). After that, by taking the superposition of \(\ket{\eta_1}\) and \(\ket{\eta_2}\), we can also introduce the following normalized state: 
\begin{eqnarray}\label{Eta-initial}
\ket{\eta} &=& \Psi_1 \ket{\eta_1} + \Psi_2 \ket{\eta_2} \nonumber \\
&=& \psi_1 \big(\Psi_1 \ket{e_1} + \Psi_2 \ket{e_2}\big) + \psi_2 \big(\Psi_1 \ket{e_3} + \Psi_2 \ket{e_4}\big). \quad
\end{eqnarray}
Hence, the expression for Eq.~\eqref{Explicit-e-rho} becomes
\begin{eqnarray}\label{Erho'Eta}
e_{(\rho')} = \ket{\eta}\bra{\eta}.
\end{eqnarray}
We have addressed the first component of the expression on the right-hand side of Eq.~\eqref{Rho'RhotimesTau}, that is, \(e_{(\rho')}\). To obtain the explicit form of the second component of the expression on the right-hand side of Eq.~\eqref{Rho'RhotimesTau}, that is, \(o_{(\rho')}\), a simple substitution in Eq.~\eqref{Explicit-e-rho} suffices, where we replace \(e\) with \(o\) and \(\Psi\) with \(\Phi\). We can proceed then by employing analogous steps. Namely, we first introduce \(\ket{\zeta_1} = \psi_1 \ket{o_1} + \psi_2 \ket{o_3}\) and \(\ket{\zeta_2} = \psi_1 \ket{o_2} + \psi_2 \ket{o_4}\), and then take the superposition of states \(\ket{\zeta_1}\) and \(\ket{\zeta_2}\). In this way, we arrive at the following normalized state: 
\begin{eqnarray}\label{Zeta-initial}
\ket{\zeta} &=& \Phi_1 \ket{\zeta_1} + \Phi_2 \ket{\zeta_2} \nonumber \\
&=& \psi_1 \big(\Phi_1 \ket{o_1} + \Phi_2 \ket{o_2}\big) + \psi_2 \big(\Phi_1 \ket{o_3} + \Phi_2 \ket{o_4}\big). \quad
\end{eqnarray}
By considering the state \(\ket{\zeta}\) and utilizing the explicit form of Eq.~\eqref{Explicit-e-rho} corresponding to \(o_{(\rho')}\), we arrive at the final expression: 
\begin{eqnarray}\label{Orho'Zeta}
o_{(\rho')} = \ket{\zeta}\bra{\zeta}.
\end{eqnarray}
As a result, the expressions for \(e_{(\rho')}\) and \(o_{(\rho')}\) in Eq.~\eqref{Rho'RhotimesTau} are derived and represented by Eqs.~\eqref{Erho'Eta} and \eqref{Orho'Zeta}, respectively. Based on the obtained results, combining Eqs.~\eqref{Eta-initial} and \eqref{Erho'Eta}, we observe the following:
\begin{eqnarray}\label{InitialEvenSectorPro-General}
\mathbb{S}_{s_1}\left(\text{tr}_{B^\prime}[\rho^\prime]\right) = \Big\{\psi_1^2\Psi_1^2, \; \psi_1^2\Psi_2^2, \; \psi_2^2\Psi_1^2, \;  \psi_2^2\Psi_2^2\Big\}.
\end{eqnarray}
This probability distribution corresponds specifically to the even sector of the initial state. In the same manner, combining Eqs.~\eqref{Zeta-initial} and \eqref{Orho'Zeta}, we observe the following:
\begin{eqnarray}\label{InitialOddSectorPro-General}
\mathbb{S}_{s_2}\left(\text{tr}_{B^\prime}[\rho^\prime]\right) = \Big\{\psi_1^2\Phi_1^2, \; \psi_1^2\Phi_2^2, \; \psi_2^2\Phi_1^2, \; \psi_2^2\Phi_2^2\Big\}.
\end{eqnarray}
These probabilities correspond to the odd sector. Thus, we have clarified the required knowledge about \(\rho \wedge \tau\).

We can apply the same design approach to the final state. The expression of the target state, augmented by the wedge product of \(\sigma\) and \(\tau\), is given by
\begin{eqnarray}\label{sigma'SigmatimesTau}
\sigma' = \sigma \wedge \tau = \frac{1}{2}\big(e_{(\sigma')}  + o_{(\sigma')}\big).   
\end{eqnarray}
Again, following a series of mathematical calculations, we deduce that 
\begin{widetext}
\begin{eqnarray}\label{Explicit-e-sigma}
e_{(\sigma')}
&=& {\Phi_1^2} \bigg[\Big(\phi_1 \ket{e_5} + \phi_2 \ket{e_7}\Big)\Big(\phi_1 \bra{e_5} + \phi_2 \bra{e_7}\Big)\bigg] + {\Phi_2^2} \bigg[\Big(\phi_1 \ket{e_6} + \phi_2 \ket{e_8}\Big)\Big(\phi_1 \bra{e_6} + \phi_2 \bra{e_8}\Big)\bigg]  \nonumber \\
&& + {\Phi_1\Phi_2} \bigg[\phi_1\phi_2\Big(\ket{e_5}\bra{e_8} + \ket{e_8}\bra{e_5} + \ket{e_6}\bra{e_7} + \ket{e_7}\bra{e_6}\Big) + \phi_1^2 \Big(\ket{e_5}\bra{e_6} +\ket{e_6}\bra{e_5}\Big) + \phi_2^2 \Big(\ket{e_7}\bra{e_8} +\ket{e_8}\bra{e_7}\Big)\bigg]. \; \nonumber \\
\end{eqnarray}
\end{widetext}
Further simplification of the expression in Eq.~\eqref{Explicit-e-sigma} is possible by introducing  \(\ket{\omega_1} = \phi_1 \ket{e_5} + \phi_2 \ket{e_7}\) and 
\(\ket{\omega_2} = \phi_1 \ket{e_6} + \phi_2 \ket{e_8}\). After that, by taking the superposition of \(\ket{\omega_1}\) and \(\ket{\omega_2}\), we arrive at the following normalized state:  
\begin{eqnarray}\label{Omega-Final}
\ket{\omega} &=& \Phi_1 \ket{\omega_1} + \Phi_2 \ket{\omega_2} \nonumber \\
&=& \phi_1 \big(\Phi_1 \ket{e_5} + \Phi_2 \ket{e_6}\big) + \phi_2 \big(\Phi_1 \ket{e_7} + \Phi_2 \ket{e_8}\big). \quad
\end{eqnarray}
Hence, the expression for Eq.~\eqref{Explicit-e-sigma} becomes 
\begin{eqnarray}\label{Esigma'Omega}
e_{(\sigma')} = \ket{\omega}\bra{\omega}.
\end{eqnarray}
To obtain the explicit form of \(o_{(\sigma')}\), a simple substitution in Eq.~\eqref{Explicit-e-sigma} suffices, where we replace \(e\) with \(o\) and \(\Phi\) with \(\Psi\). We can follow the usual procedure by introducing analogous steps. Specifically, we define  \(\ket{\xi_1} = \phi_1 \ket{o_5} + \phi_2 \ket{o_7}\) and \(\ket{\xi_2} = \phi_1 \ket{o_6} + \phi_2 \ket{o_8}\), and then superpose \(\ket{\xi_1}\) and \(\ket{\xi_2}\) to obtain the normalized state:
\begin{eqnarray}\label{Xi-Final}
\ket{\xi} &=& \Psi_1 \ket{\xi_1} + \Psi_2 \ket{\xi_2} \nonumber \\
&=& \phi_1 \big(\Psi_1 \ket{o_5} + \Psi_2 \ket{o_6}\big) + \phi_2 \big(\Psi_1 \ket{o_7} + \Psi_2 \ket{o_8}\big). \quad
\end{eqnarray}
Taking into account the state \(\ket{\xi}\) and employing the specific form of Eq.~\eqref{Explicit-e-sigma} corresponding to \(o_{(\sigma')}\), we obtain the resulting expression:
\begin{eqnarray}\label{Osigma'Xi}
o_{(\sigma')} = \ket{\xi}\bra{\xi}.
\end{eqnarray}
As a result, the expressions for \(e_{(\sigma')}\) and \(o_{(\sigma')}\) in Eq.~\eqref{sigma'SigmatimesTau} are derived and represented by Eqs.~\eqref{Esigma'Omega} and \eqref{Osigma'Xi}, respectively. Therefore, based on the obtained results, combining Eqs.~\eqref{Omega-Final} and \eqref{Esigma'Omega}, we observe the following:
\begin{eqnarray}\label{FinalEvenSectorPro-General}
\mathbb{S}_{s_1}\left(\text{tr}_{B^\prime}[\sigma^\prime]\right) = \Big\{\phi_1^2\Phi_1^2, \; \phi_1^2\Phi_2^2, \; \phi_2^2\Phi_1^2, \;  \phi_2^2\Phi_2^2\Big\}.
\end{eqnarray}
These probabilities correspond to the even sector. Similarly, combining Eqs.~\eqref{Xi-Final} and \eqref{Osigma'Xi}, we observe the following:
\begin{eqnarray}\label{FinalOddSectorPro-General}
\mathbb{S}_{s_2}\left(\text{tr}_{B^\prime}[\sigma^\prime]\right) = \Big\{\phi_1^2\Psi_1^2, \; \phi_1^2\Psi_2^2, \; \phi_2^2\Psi_1^2, \; \phi_2^2\Psi_2^2\Big\}.
\end{eqnarray}
These probabilities correspond to the odd sector. Thus, we have clarified the required knowledge about \(\sigma'\). With this, we have concluded another phase of our analysis. We now possess comprehensive knowledge regarding the explicit representations of both the initial and target states. Formally, we have 
\begin{eqnarray}\label{Even-E}
\rho' = \frac{1}{2}\big({e_{(\rho')}} + {o_{(\rho')}}\big) \overset{\mathbb{E}({e_{\rho'} \rightarrow  e_{\sigma'}})}{\longrightarrow} 
\varrho = \frac{1}{2}\big({e_{(\sigma')}} + {o_{(\rho')}}\big),  
\end{eqnarray}
\begin{eqnarray}\label{Odd-O}
\varrho = \frac{1}{2}\big({e_{(\sigma')}} + {o_{(\rho')}}\big) \overset{\mathbb{O}({o_{\rho'} \rightarrow o_{\sigma'}})}{\longrightarrow} 
\sigma' = \frac{1}{2}\big({e_{(\sigma')}} + {o_{(\sigma')}}\big).  
\end{eqnarray}
The first transformation \eqref{Even-E} specifically involves converting the pure state \(\ket{\eta}\) given in Eq.~\eqref{Eta-initial} into the pure state \(\ket{\omega}\) given in Eq.~\eqref{Omega-Final}, while the second transformation \eqref{Odd-O} is focused on converting the pure state \(\ket{\zeta}\) given in Eq.~\eqref{Zeta-initial} into the pure state \(\ket{\xi}\) given in Eq.~\eqref{Xi-Final}.

\subsection{Explicit example for catalytic transformations}\label{SubSec::Example2}

We now turn our attention to the given example in Sec.~\ref{SubSec::Cat-Transform}. Considering Example 2,  when we rewrite Eqs.~\eqref{InitialEvenSectorPro-General}  and \eqref{FinalEvenSectorPro-General}, it becomes evident that they correspond to equations Eqs.~\eqref{FinalSectors-initial} and \eqref{FinalSectors-target}, respectively, as expected. Then, by reordering the probability distributions \eqref{FinalSectors-initial} and \eqref{FinalSectors-target} in a non-increasing order, we obtain the following sequences:
\begin{eqnarray}\label{FinalSectors-initial-EvenOrdered}
\mathbb{S}_{s_1}\left(\text{tr}_{B^\prime}[\rho^\prime]\right) = \big\{0.63,  0.21,  0.12, 0.04\big\},
\end{eqnarray}
\begin{eqnarray}\label{FinalSectors-target-EvenOrdered}
\mathbb{S}_{s_1}\left(\text{tr}_{B^\prime}[\sigma^\prime]\right) = \big\{0.6825, 0.2275, 0.0675, 0.0225\big\}.
\end{eqnarray}
Here it is obvious that \(\mathbb{S}_{s_1}\big(\text{tr}_{B^\prime}[\rho^\prime]\big) \prec \mathbb{S}_{s_1}\big(\text{tr}_{B^\prime}[\sigma^\prime]\big)\). The operation denoted as \(\mathbb{E}({e_{\rho'} \rightarrow e_{\sigma'}})\) in Eq.~\eqref{First-Step-varRho2}, achieves the conversion of the even sector of the initial state \(\rho'\) into the even sector of the target state \(\sigma'\) with unit probability, As we mentioned before, the application of different permutation sets is necessary, based on the relationships among coefficients in the initial and target states. The precise identification of these sets is integral for computing probabilities and, as a result, establishing local operators. For the specific case where \(0.21 \leq 0.2275\) and \(0.12 \geq 0.0675\) with \(0.21+0.12 \geq 0.2275+0.0675\), we make use of Table \ref{TablePermutations4D4th1} provided in the Supplemental Material. With the assistance of Table \ref{TablePermutations4D4th1}, it becomes possible to construct the operators \(\hat{\mathcal{E}}_{A'}^{3}\) and \(\hat{\mathcal{E}}_{B'}^{3}\), for instance, as follows:
\begin{widetext}
\begin{eqnarray}
\hat{\mathcal{E}}_{A'}^{3} &=& \sqrt{\mathfrak{p}_3} \left[\sqrt{\frac{68.25}{63}} \; \ket{e_{A',5}}\bra{e_{A',1}} + \sqrt{\frac{6.75}{21}} \; \ket{e_{A',7}}\bra{e_{A',2}} + \sqrt{\frac{22.75}{12}} \; \ket{e_{A',6}}\bra{e_{A',3}} + \sqrt{\frac{2.25}{4}} \; \ket{e_{A',8}}\bra{e_{A',4}}\right] \nonumber \\
& & \oplus 
\sqrt{\mathfrak{p}_3} \sum_{i=1}^{4}\ket{o_{A',i}}\bra{o_{A',i}},
\end{eqnarray}
\begin{eqnarray}
\hat{\mathcal{E}}_{B'}^{3} = 
\Big[\ket{e_{B',5}}\bra{e_{B',1}} + \ket{e_{B',7}}\bra{e_{B',2}} + \ket{e_{B',6}}\bra{e_{B',3}} + \ket{e_{B',8}}\bra{e_{B',4}}\Big] \oplus 
\sum_{i=1}^{4}\ket{o_{B',i}}\bra{o_{B',i}},
\end{eqnarray}
\end{widetext}
where \({\mathfrak{p}_3} = 0.109375\). By referring to Table \ref{TablePermutations4D4th1}, we can readily construct the remaining local operators \(\hat{\mathcal{E}}_{A'}^{k}\) and \(\hat{\mathcal{E}}_{B'}^{k}\) for \(k=1, 2, 4\). As a result, we obtain an intermediate state \(\varrho=(e_{(\sigma')}+o_{(\rho')})/{2}\). We then proceed with the transformation of the odd sector. The operation denoted as \(\mathbb{O}({o_{\rho'} \rightarrow o_{\sigma'}})\) in Eq.~\eqref{Second-Step-varRho2}, achieves the conversion of the odd sector of the intermediate state \(\varrho\) (i.e., \(\rho'\)) into the odd sector of the target state \(\sigma'\) with unit probability. It has been noted that the elements of the odd sectors are identical to those given in Eqs.\eqref{FinalSectors-initial-EvenOrdered} and \eqref{FinalSectors-target-EvenOrdered}. Therefore, we can once again refer to Table \ref{TablePermutations4D4th1} from the Supplemental Material. By utilizing Table \ref{TablePermutations4D4th1}, we can construct the operators \(\hat{\mathcal{O}}_{A'}^{2}\) and \(\hat{\mathcal{O}}_{B'}^{2}\), for instance, in such a way that
\begin{widetext}
\begin{eqnarray}
\hat{\mathcal{O}}_{A'}^{2} &=& \sqrt{\mathfrak{q}_2} \sum_{i=5}^{8}\ket{e_{A',i}}\bra{e_{A',i}} \nonumber \\
& & 
\oplus \sqrt{\mathfrak{q}_2} \left[\sqrt{\frac{2.25}{63}} \; \ket{o_{A',8}}\bra{o_{A',1}} + \sqrt{\frac{22.75}{21}} \; \ket{o_{A',6}}\bra{o_{A',2}} + \sqrt{\frac{6.75}{12}} \; \ket{o_{A',7}}\bra{o_{A',3}} + \sqrt{\frac{68.25}{4}} \; \ket{o_{A',5}}\bra{o_{A',4}}\right],
\end{eqnarray}
\begin{eqnarray}
\hat{\mathcal{O}}_{B'}^{2} = \sum_{i=5}^{8}\ket{e_{B',i}}\bra{e_{B',i}}
\oplus 
\Big[\ket{o_{B',8}}\bra{o_{B',1}} + \ket{o_{B',6}}\bra{o_{B',2}} + \ket{o_{B',7}}\bra{o_{B',3}} + \ket{o_{B',5}}\bra{o_{B',4}}\Big],
\end{eqnarray}
\end{widetext}
where \({\mathfrak{q}_2} = 0.026515\). Moreover, by referencing Table \ref{TablePermutations4D4th1}, we can construct other local operators  \(\hat{\mathcal{O}}_{A'}^{k}\) and \(\hat{\mathcal{O}}_{B'}^{k}\)  for \(k=1, 3, 4\) with ease. As a result, we obtain the target state \(\sigma'\). In conclusion, the desired transformations can be effectively achieved by utilizing the established structure and referring to the tables provided in the Supplemental Material (Sec.~\ref{Supp:PermAndProb}). These carefully constructed tables, along with the systematic framework, serve as invaluable resources for accomplishing the desired transformations in our analysis.

\section{Discussions}\label{Sec:Conc}

In the present paper, we have aimed at providing a convenient framework for the investigation of entanglement transformations in fermionic mode systems under local SSRs. To this end, we proposed a necessary and sufficient set of conditions in Algorithm~\ref{alg:algoritma}. We focused on a two-orbital system that perfectly exemplifies the differences between distinguishable and indistinguishable bipartite quantum systems. Our framework also enabled us to demonstrate how the local parity of orbital subsystems can be changed with the help of an ancillary orbital system called a \textit{catalyst}.

The same framework can be applied to a variety of problems beyond the example given here. For instance, many of the orbital correlations in quantum chemical systems become inaccessible in the presence of local SSRs~\cite{2021_JCTC_DMRG_SSR, OrbitalDiscord}. We are planning to investigate the possibility of activating such inaccessible orbital correlations by using the notions of catalytic state transformations and majorization.

Besides this, the idea of a portion of superposition that is not useful for free operations and should be eliminated during the transformation is not restricted only to the fermionic mode systems; it may also contribute to the resource theory of quantum correlations in distinguishable quantum systems. As an example, the state transformations between W-type and GHZ-type tripartite entanglement classes require similar constraints to the state transformation discussed in Example 2 (see Sec.~\ref{Sec::Catalyt}). A natural question to ask at this point is that would it be possible to convert a W state to a GHZ state or \textit{vice versa} with the help of a catalyst.

Another notion of distinguishable system nonlocality to pursue future work by our methodology is the bipartite quantum correlations beyond entanglement, namely the quantum discord \cite{Ollivier2001QD}, which can be possessed even in some separable states. LOCC, or more generally the so-called separable operations can raise the amount of quantum discord distributed in a bipartite system. Hence, such operations alone are not sufficient to describe the free operations in the theory of discord. In this regard, we believe the framework developed here may pave the way for a more detailed understanding of the free operations in the resource theories of both multipartite quantum entanglement and bipartite quantum discord.

\begin{acknowledgments}	
We acknowledge financial support from the Scientific and Technological Research Council of Turkey (TÜBİTAK) under Grant No.~120F089.
\end{acknowledgments}

\bibliographystyle{apsrev4-2}
%


\pagebreak

\widetext
\begin{center}
\textbf{Supplemental Material:} 
\vskip 3mm
\textbf{Manipulating fermionic mode entanglement in the presence of superselection rules}
\end{center}

In this Supplemental Material we provide further details for the manuscript ``\emph{Manipulating fermionic mode entanglement in the presence of superselection rules}'', including the algorithm for the biparite entanglement transformation \(\rho \mapsto \Lambda^{\text{P}}(\rho)=\sigma\) and the integral elements of the local fermionic quantum operators that are characterized by their explicit forms and detailed in Eqs.~\eqref{Local-Op-Epsilon-A}, \eqref{Local-Op-Epsilon-B}, \eqref{Local-Op-O-A}, and \eqref{Local-Op-O-B} in the main text. 

\begin{algorithm}[h] 
\label{alg:algoritma}

\eIf{$\chi_{\sigma}=0$}
{

    $n \gets 0$\;
    $j \gets 0$\;
    \While{$j \leq \textbf{dim}(\rho)$}
    {
    \If{\(P_{s_j}=Q_{s_j}\)}
    {
        $n \gets n+1$
    }
    $j \gets j+1$
    }
    \eIf{$n = \textbf{dim}(\rho)$}
    {
        $n \gets 0$\;
        $j \gets 0$\;
        \While{$j \leq \textbf{dim}(\rho)$}
        {
        \If{\(\mathbb{S}_{s_j}(\rho_B^{\text{P}}) \prec \mathbb{S}_{s_j}(\sigma_B^{\text{P}})\)}
        {
            $n \gets n+1$
        }
        $j \gets j+1$
        }
        \eIf{$n = \textbf{dim}(\rho)$}
        {
        The transformation is possible.
        }
        {
        The transformation is not possible.
        }
    }
    {
    The transformation is not possible.
    }

}
{
The transformation is not possible.
}

\caption{Algorithm for the transformation of bipartite mode entanglement under Q-SSR}
\end{algorithm}

\section{Integral elements of the operations} \label{Supp:PermAndProb}

This section plays a pivotal role in constructing operations \(\mathbb{E}({e_{\rho'} \rightarrow e_{\sigma'}})\) and \(\mathbb{O}({o_{\rho'} \rightarrow o_{\sigma'}})\). To effectively utilize the presented information, we provide a concise guide for interpreting the content. The collection of writings provided here is sourced from Ref.~\cite{TorunCOHTR18} and reorganized accordingly.

Let us suppose that we have two probability distributions, denoted as 
\begin{eqnarray}\label{XandY-dist}
{x} = \left\{{x}_{1}^{\downarrow}, {x}_{2}^{\downarrow}, {x}_{3}^{\downarrow}, {x}_{4}^{\downarrow}\right\}, \quad
{y} = \left\{{y}_{1}^{\downarrow}, {y}_{2}^{\downarrow}, {y}_{3}^{\downarrow}, {y}_{4}^{\downarrow}\right\},
\end{eqnarray} 
where the elements are arranged in non-increasing order, satisfying \(\sum_{i=1}^{4}{x}_{1}^{\downarrow}=1\) and \(\sum_{i=1}^{4}{y}_{1}^{\downarrow}=1\). Moving forward, we omit the symbol ``\(\downarrow\)'' and instead use the notations \(\{x_1, x_2, x_3, x_4\}\) and \(\{y_1, y_2, y_3, y_4\}\). We can express the majorization relationship between these distributions as \({x} \prec {y} \), indicating that \({x} \) is majorized by \({y}\) (i.e., \(\sum_{i=1}^{k} x_{i} \leq \sum_{i=1}^{k} y_{i}\) for all \(k=1, 2, 3, 4\)). In this scenario, we can confidently assert that \({x} _{1} \leq {y} _{1}\) and \({x} _{4} \geq {y} _{4}\) due to the majorization condition. However, it is important to note that there is no singular relationship between \({x} _{2}\) and \({y} _{2}\) or between \({x} _{3}\) and \({y} _{3}\). In other words, \({x} _{2}\)  can either be greater than or equal to \({y} _{2}\) (\({x} _{2} \geq {y} _{2}\)) as well as greater than or equal to \({y} _{2}\) (\({x} _{2} \leq {y} _{2}\)). Similarly, this variability applies to \({x}_{3}\)  and \({y}_{3}\). This highlights the necessity of considering these diverse situations as we advance in our analysis.

There exist four possible scenarios to consider: 
\begin{equation}
\begin{cases}
      \text{(C1)}: & {x}_{2} \geq {y}_{2} \quad {\&} \quad {x}_{3} \geq {y}_{3}, \\
      \text{(C2)}: & {x}_{2} \geq {y}_{2} \quad {\&} \quad {x}_{3} \leq {y}_{3}, \\
      \text{(C3)}: & {x}_{2} \leq {y}_{2} \quad {\&} \quad {x}_{3} \leq {y}_{3}, \\
      \text{(C4)}: & {x}_{2} \leq {y}_{2} \quad {\&} \quad {x}_{3} \geq {y}_{3}, \quad 
{\begin{cases}
      \text{(SC1)}: \quad {x}_{2}+{x}_{3} \geq {y}_{2}+{y}_{3},  \\  \\
      \text{(SC2)}: \quad {x}_{2}+{x}_{3} \leq {y}_{2}+{y}_{3}. 
\end{cases}}
\end{cases}
\end{equation}
For each individual case, we construct a complete table encompassing the permutations \(\{\pi_{i}^{k}\}\), the function \(f_{k}(i)\) (and \(g_{k}(j)\)), and the probabilities \(\{\mathfrak{p}_k\}\) (and \(\{\mathfrak{q}_k\}\)). Tables for each case (C1), (C2), (C3), and (C4) are presented below, respectively. It is important to note that the last case (C4) consists of two separate subcases (SC1) and (SC2). The necessary explanations for each case can be found in the captions of the respective tables. Thus, the tabulated representation in Tables \ref{TablePermutations4D1st}, \ref{TablePermutations4D2nd}, \ref{TablePermutations4D3rd}, \ref{TablePermutations4D4th1}, and \ref{TablePermutations4D4th2} serves as a valuable tool for examining and discerning the different possibilities involved. By analyzing the progression of Example 2 discussed in the main text, a better understanding of how to utilize the tables can also be gained.

Now, let us take a moment to revisit and remind ourselves of the distributions for which we seek majorization relationships in the main text. It is important to refresh our understanding of these distributions before proceeding further. Hence, if we recall Eqs.~\eqref{InitialEvenSectorPro-General}, \eqref{InitialOddSectorPro-General}, \eqref{FinalEvenSectorPro-General}, and \eqref{FinalOddSectorPro-General}, we have 
\begin{eqnarray}\label{Supp-InitialDis}
\mathbb{S}_{s_1}\left(\text{tr}_{B^\prime}[\rho^\prime]\right) = \Big\{\psi_1^2\Psi_1^2, \; \psi_1^2\Psi_2^2, \; \psi_2^2\Psi_1^2, \; \psi_2^2\Psi_2^2\Big\}, \quad 
\mathbb{S}_{s_2}\left(\text{tr}_{B^\prime}[\rho^\prime]\right) = \Big\{\psi_1^2\Phi_1^2, \; \psi_1^2\Phi_2^2, \; \psi_2^2\Phi_1^2, \; \psi_2^2\Phi_2^2\Big\},
\end{eqnarray}
\begin{eqnarray}\label{Supp-FinalDis}
\mathbb{S}_{s_1}\left(\text{tr}_{B^\prime}[\sigma^\prime]\right) = \Big\{\phi_1^2\Phi_1^2, \; \phi_1^2\Phi_2^2, \; \phi_2^2\Phi_1^2, \; \phi_2^2\Phi_2^2\Big\}, \quad 
\mathbb{S}_{s_2}\left(\text{tr}_{B^\prime}[\sigma^\prime]\right) = \Big\{\phi_1^2\Psi_1^2, \; \phi_1^2\Psi_2^2, \; \phi_2^2\Psi_1^2, \; \phi_2^2\Psi_2^2\Big\}.
\end{eqnarray}
In our analysis, we consider the majorization comparisons between the pairs \(\mathbb{S}_{s_1}\big(\text{tr}_{B^\prime}[\rho^\prime]\big)\) and \(\mathbb{S}_{s_1}\big(\text{tr}_{B^\prime}[\sigma^\prime]\big)\), as well as the pairs \(\mathbb{S}_{s_2}\big(\text{tr}_{B^\prime}[\rho^\prime]\big)\) and \(\mathbb{S}_{s_2}\big(\text{tr}_{B^\prime}[\sigma^\prime]\big)\). More precisely, 
we aim to determine whether the following majorization relationships hold:
\begin{equation}
\mathbb{S}_{s_1}\big(\text{tr}_{B^\prime}[\rho^\prime]\big) \prec  \mathbb{S}_{s_1}\big(\text{tr}_{B^\prime}[\sigma^\prime]\big) 
\end{equation}
and
\begin{equation}
\mathbb{S}_{s_2}\big(\text{tr}_{B^\prime}[\rho^\prime]\big) \prec  \mathbb{S}_{s_2}\big(\text{tr}_{B^\prime}[\sigma^\prime]\big).
\end{equation}
Here, we compare the majorization relationships between the partial traces of the initial state \(\rho'\) and the target state \(\sigma'\) with respect to the subsystem \(B'\). These comparisons are performed using the spectral mappings \(\mathbb{S}_{s_1}\) and \(\mathbb{S}_{s_2}\), respectively.  Furthermore, for the sake of generality, we can assume that the distributions described in Eqs.~\eqref{Supp-InitialDis} and \eqref{Supp-FinalDis} are presented in non-increasing order. However, it is worth noting that the examples under consideration may have a mixed order, which is permissible. In such cases, the samples can be rearranged in non-increasing order through the application of suitable unitary transformations (see Example 2 in the main text). Therefore, we can conclude that the tables constructed based on the distributions provided in Eq.~\eqref{XandY-dist} are specifically designed to apply to the distributions presented in Eqs.~\eqref{Supp-InitialDis} and \eqref{Supp-FinalDis}. By making this connection, we can ensure the proper utilization of the tables for the intended distributions.


\begin{table*}[htb!] 
\caption{The table includes the permutations \(\{{\pi}_{i}^{k}\}\), the function \(f_{k}(i) = \sqrt{{y_a}/{x_b}}\) denoted as \(\varkappa_{ab}\) for \(a, b = 1, 2, 3, 4\), and the corresponding probabilities \(\{\mathfrak{p}_k\}\). These integral elements contribute to the formulation of the local operators \(\{\hat{\mathcal{E}}_{A'}^{k}\}\) and \(\{\hat{\mathcal{E}}_{B'}^{k}\}\) given by Eqs.~\eqref{Local-Op-Epsilon-A} and \eqref{Local-Op-Epsilon-B}, respectively. In addition, the formulation of the local operators \(\{\hat{\mathcal{O}}_{A'}^{k}\}\) and \(\{\hat{\mathcal{O}}_{B'}^{k}\}\) given by Eqs.~\eqref{Local-Op-O-A} and \eqref{Local-Op-O-B}, respectively, follows a similar construction, where the substitution of variables is applied. Specifically, replacing \(\{{\pi}_{i}^{k}\}\) with \(\{{\pi}_{j}^{k}\}\), \(f_k(i)\) with \(g_k(j)\), and \(\mathfrak{p}\) with \(\mathfrak{q}\) yields the corresponding elements. Importantly, the entries in the table are applicable to the scenario where \({x}_{2} \geq {y}_{2}\) and \({x}_{3} \geq {y}_{3}\). We have omitted the subscripts \(A'\) and \(B'\) in the representation of states for the sake of simplicity.}
\label{TablePermutations4D1st}
\centering
\begin{ruledtabular}
\begin{tabular}{c c c}
\multicolumn{1}{c}{Permutations for the case \({x}_{2} \geq {y}_{2}\) \(\&\) \({x}_{3} \geq {y}_{3}\)} &
\multicolumn{1}{c}{The function \(f_{k}(i)\)} &
\multicolumn{1}{c}{The probabilities}  \Tstrut\Bstrut \\ [1ex] 
\cline{1-1}  \cline{2-2}  \cline{3-3}  \\ [-1ex]
  \({\pi}_{1}^{1}\ket{e_1} = \ket{e_5}\) ; \; \({\pi}_{2}^{1}\ket{e_2} = \ket{e_6}\) ; \;
\({\pi}_{3}^{1}\ket{e_3} = \ket{e_7}\) ; \; \({\pi}_{4}^{1}\ket{e_4} = \ket{e_8}\)  & \(\varkappa_{11}; \; \varkappa_{22}; \; \varkappa_{33}; \; \varkappa_{44}\) &
\(\mathfrak{p}_1 = 1-\sum_{k=2}^{4} \mathfrak{p}_k\)  \\   [2ex]
  \({\pi}_{1}^{2}\ket{e_1} = \ket{e_8}\) ; \; \({\pi}_{2}^{2}\ket{e_2} = \ket{e_6}\) ; \; 
\({\pi}_{3}^{2}\ket{e_3} = \ket{e_7}\) ; \; \({\pi}_{4}^{2}\ket{e_4} = \ket{e_5}\)  & \(\varkappa_{41}; \; \varkappa_{22}; \; \varkappa_{33}; \; \varkappa_{14}\) &
\(\mathfrak{p}_2={({x}_{4} - {y}_{4})}/{({y}_{1} - {y}_{4})}\)  \\   [2ex]
  \({\pi}_{1}^{3}\ket{e_1} = \ket{e_6}\) ; \; \({\pi}_{2}^{3}\ket{e_2} = \ket{e_5}\) ; \; 
\({\pi}_{3}^{3}\ket{e_3} = \ket{e_7}\) ; \; \({\pi}_{4}^{3}\ket{e_4} = \ket{e_8}\)  & \(\varkappa_{21}; \; \varkappa_{12}; \; \varkappa_{33}; \; \varkappa_{44}\) &
\(\mathfrak{p}_3={({x}_{2} - {y}_{2})}/{({y}_{1} - {y}_{2})}\)  \\   [2ex]
  \({\pi}_{1}^{4}\ket{e_1} = \ket{e_7}\) ; \; \({\pi}_{2}^{4}\ket{e_2} = \ket{e_6}\) ; \; 
\({\pi}_{3}^{4}\ket{e_3} = \ket{e_5}\) ; \; \({\pi}_{4}^{4}\ket{e_4} = \ket{e_8}\)  & \(\varkappa_{31}; \; \varkappa_{22}; \; \varkappa_{13}; \; \varkappa_{44}\) & 
\(\mathfrak{p}_4={({x}_{3} - {y}_{3})}/{({y}_{1} - {y}_{3})}\)  \\   [2ex]
\end{tabular}
\end{ruledtabular}
\end{table*}


\begin{table*}[htb!] 
\caption{The table includes the permutations \(\{{\pi}_{i}^{k}\}\), the function \(f_{k}(i) = \sqrt{{y_a}/{x_b}}\) denoted as \(\varkappa_{ab}\) for \(a, b = 1, 2, 3, 4\), and the corresponding probabilities \(\{\mathfrak{p}_k\}\). These integral elements contribute to the formulation of the local operators \(\{\hat{\mathcal{E}}_{A'}^{k}\}\) and \(\{\hat{\mathcal{E}}_{B'}^{k}\}\) given by Eqs.~\eqref{Local-Op-Epsilon-A} and \eqref{Local-Op-Epsilon-B}, respectively. In addition, the formulation of the local operators \(\{\hat{\mathcal{O}}_{A'}^{k}\}\) and \(\{\hat{\mathcal{O}}_{B'}^{k}\}\) given by Eqs.~\eqref{Local-Op-O-A} and \eqref{Local-Op-O-B}, respectively, follows a similar construction, where the substitution of variables is applied. Specifically, replacing \(\{{\pi}_{i}^{k}\}\) with \(\{{\pi}_{j}^{k}\}\), \(f_k(i)\) with \(g_k(j)\), and \(\mathfrak{p}\) with \(\mathfrak{q}\) yields the corresponding elements. Importantly, the entries in the table are applicable to the scenario where \({x}_{2} \geq {y}_{2}\) and \({x}_{3} \leq {y}_{3}\). We have omitted the subscripts \(A'\) and \(B'\) in the representation of states for the sake of simplicity.}
\label{TablePermutations4D2nd}
\centering
\begin{ruledtabular}
\begin{tabular}{c c c}
\multicolumn{1}{c}{Permutations for the case \({x}_{2} \geq {y}_{2}\) \(\&\) \({x}_{3} \leq {y}_{3}\)} & 
\multicolumn{1}{c}{The function \(f_{k}(i)\)} &
\multicolumn{1}{c}{The probabilities}\Tstrut\Bstrut \\ [1ex] 
\cline{1-1}  \cline{2-2}  \cline{3-3} \\ [-1ex]
\({\pi}_{1}^{1}\ket{e_1} = \ket{e_5}\) ; \; 
\({\pi}_{2}^{1}\ket{e_2} = \ket{e_6}\) ; \;  
\({\pi}_{3}^{1}\ket{e_3} = \ket{e_7}\) ; \; 
\({\pi}_{4}^{1}\ket{e_4} = \ket{e_8}\) & \(\varkappa_{11}; \; \varkappa_{22}; \; \varkappa_{33}; \; \varkappa_{44}\) &  
\(\mathfrak{p}_1 = 1-\sum_{k=2}^{4} \mathfrak{p}_k\)  \\   [2ex]
  \({\pi}_{1}^{2}\ket{e_1} = \ket{e_8}\) ; \; \({\pi}_{2}^{2}\ket{e_2} = \ket{e_6}\) ; \; 
\({\pi}_{3}^{2}\ket{e_3} = \ket{e_7}\) ; \; \({\pi}_{4}^{2}\ket{e_4} = \ket{e_5}\) & \(\varkappa_{41}; \; \varkappa_{22}; \; \varkappa_{33}; \; \varkappa_{14}\) &
\(\mathfrak{p}_2={({x}_{3}+{x}_{4} - {y}_{3}-{y}_{4})}/{({y}_{1} - {y}_{4})}\)  \\   [2ex]
  \({\pi}_{1}^{3}\ket{e_1} = \ket{e_6}\) ; \; \({\pi}_{2}^{3}\ket{e_2} = \ket{e_5}\) ; \; 
\({\pi}_{3}^{3}\ket{e_3} = \ket{e_7}\) ; \; \({\pi}_{4}^{3}\ket{e_4} = \ket{e_8}\) & \(\varkappa_{21}; \; \varkappa_{12}; \; \varkappa_{33}; \; \varkappa_{44}\) &  
\(\mathfrak{p}_3={({x}_{2} - {y}_{2})}/{({y}_{1} - {y}_{2})}\)  \\   [2ex]
  \({\pi}_{1}^{4}\ket{e_1} = \ket{e_5}\) ; \; \({\pi}_{2}^{4}\ket{e_2} = \ket{e_6}\) ; \; 
\({\pi}_{3}^{4}\ket{e_3} = \ket{e_8}\) ; \; \({\pi}_{4}^{4}\ket{e_4} = \ket{e_7}\) & \(\varkappa_{11}; \; \varkappa_{22}; \; \varkappa_{43}; \; \varkappa_{34}\) & \(\mathfrak{p}_4={({y}_{3} - {x}_{3})}/{({y}_{3} - {y}_{4})}\)  \\   [2ex]
\end{tabular}
\end{ruledtabular}
\end{table*}


\begin{table*}[htb!] 
\caption{The table includes the permutations \(\{{\pi}_{i}^{k}\}\), the function \(f_{k}(i) = \sqrt{{y_a}/{x_b}}\) denoted as \(\varkappa_{ab}\) for \(a, b = 1, 2, 3, 4\), and the corresponding probabilities \(\{\mathfrak{p}_k\}\). These integral elements contribute to the formulation of the local operators \(\{\hat{\mathcal{E}}_{A'}^{k}\}\) and \(\{\hat{\mathcal{E}}_{B'}^{k}\}\) given by Eqs.~\eqref{Local-Op-Epsilon-A} and \eqref{Local-Op-Epsilon-B}, respectively. In addition, the formulation of the local operators \(\{\hat{\mathcal{O}}_{A'}^{k}\}\) and \(\{\hat{\mathcal{O}}_{B'}^{k}\}\) given by Eqs.~\eqref{Local-Op-O-A} and \eqref{Local-Op-O-B}, respectively, follows a similar construction, where the substitution of variables is applied. Specifically, replacing \(\{{\pi}_{i}^{k}\}\) with \(\{{\pi}_{j}^{k}\}\), \(f_k(i)\) with \(g_k(j)\), and \(\mathfrak{p}\) with \(\mathfrak{q}\) yields the corresponding elements. Importantly, the entries in the table are applicable to the scenario where \({x}_{2} \leq {y}_{2}\) and \({x}_{3} \leq {y}_{3}\). We have omitted the subscripts \(A'\) and \(B'\) in the representation of states for the sake of simplicity.}
\label{TablePermutations4D3rd}
\centering
\begin{ruledtabular}
\begin{tabular}{c c c}
\multicolumn{1}{c}{Permutations for the case \({x}_{2} \leq {y}_{2}\) \(\&\) \({x}_{3} \leq {y}_{3}\)} &
\multicolumn{1}{c}{The function \(f_{k}(i)\)} &
\multicolumn{1}{c}{The probabilities}\Tstrut\Bstrut \\ [1ex] 
\cline{1-1}  \cline{2-2}  \cline{3-3}  \\ [-1ex]
  \({\pi}_{1}^{1}\ket{e_1} = \ket{e_5}\) ; \; \({\pi}_{2}^{1}\ket{e_2} = \ket{e_6}\) ; \;  
\({\pi}_{3}^{1}\ket{e_3} = \ket{e_7}\) ; \; \({\pi}_{4}^{1}\ket{e_4} = \ket{e_8}\) & \(\varkappa_{11}; \; \varkappa_{22}; \; \varkappa_{33}; \; \varkappa_{44}\) &  
\(\mathfrak{p}_1 = 1-\sum_{k=2}^{4} \mathfrak{p}_k\)  \\   [2ex]
   \({\pi}_{1}^{2}\ket{e_1} = \ket{e_8}\) ; \; \({\pi}_{2}^{2}\ket{e_2} = \ket{e_6}\) ; \; 
\({\pi}_{3}^{2}\ket{e_3} = \ket{e_7}\) ; \; \({\pi}_{4}^{2}\ket{e_4} = \ket{e_5}\) & \(\varkappa_{41}; \; \varkappa_{22}; \; \varkappa_{33}; \; \varkappa_{14}\) &
\(\mathfrak{p}_2={({y}_{1}-{x}_{1})}/{({y}_{1} - {y}_{4})}\)  \\   [2ex]
    \({\pi}_{1}^{3}\ket{e_1} = \ket{e_5}\) ; \; \({\pi}_{2}^{3}\ket{e_2} = \ket{e_8}\) ; \; 
\({\pi}_{3}^{3}\ket{e_3} = \ket{e_7}\) ; \; \({\pi}_{4}^{3}\ket{e_4} = \ket{e_6}\) & \(\varkappa_{11}; \; \varkappa_{42}; \; \varkappa_{33}; \; \varkappa_{24}\) &  
\(\mathfrak{p}_3={({y}_{2} - {x}_{2})}/{({y}_{2} - {y}_{4})}\)  \\   [2ex]
    \({\pi}_{1}^{4}\ket{e_1} = \ket{e_5}\) ; \; \({\pi}_{2}^{4}\ket{e_2} = \ket{e_6}\) ; \; 
\({\pi}_{3}^{4}\ket{e_3} = \ket{e_8}\) ; \; \({\pi}_{4}^{4}\ket{e_4} = \ket{e_7}\) & \(\varkappa_{11}; \; \varkappa_{22}; \; \varkappa_{43}; \; \varkappa_{34}\) & \(\mathfrak{p}_4={({y}_{3} - {x}_{3})}/{({y}_{3} - {y}_{4})}\)  \\   [2ex]
\end{tabular}
\end{ruledtabular}
\end{table*}


\begin{table*}[htb!] 
\caption{The table includes the permutations \(\{{\pi}_{i}^{k}\}\), the function \(f_{k}(i) = \sqrt{{y_a}/{x_b}}\) denoted as \(\varkappa_{ab}\) for \(a, b = 1, 2, 3, 4\), and the corresponding probabilities \(\{\mathfrak{p}_k\}\). These integral elements contribute to the formulation of the local operators \(\{\hat{\mathcal{E}}_{A'}^{k}\}\) and \(\{\hat{\mathcal{E}}_{B'}^{k}\}\) given by Eqs.~\eqref{Local-Op-Epsilon-A} and \eqref{Local-Op-Epsilon-B}, respectively. In addition, the formulation of the local operators \(\{\hat{\mathcal{O}}_{A'}^{k}\}\) and \(\{\hat{\mathcal{O}}_{B'}^{k}\}\) given by Eqs.~\eqref{Local-Op-O-A} and \eqref{Local-Op-O-B}, respectively, follows a similar construction, where the substitution of variables is applied. Specifically, replacing \(\{{\pi}_{i}^{k}\}\) with \(\{{\pi}_{j}^{k}\}\), \(f_k(i)\) with \(g_k(j)\), and \(\mathfrak{p}\) with \(\mathfrak{q}\) yields the corresponding elements. Importantly, the entries in the table are applicable to the scenario where \({x}_{2} \leq {y}_{2}\) and \({x}_{3} \geq {y}_{3}\) with \({x}_{2}+{x}_{3} \geq {y}_{2}+{y}_{3}\). We have omitted the subscripts \(A'\) and \(B'\) in the representation of states for the sake of simplicity.}
\label{TablePermutations4D4th1}
\centering
\begin{ruledtabular}
\begin{tabular}{c c c}
\multicolumn{1}{c}{Permutations for the case \({x}_{2} \leq {y}_{2}\) \(\&\) \({x}_{3} \geq {y}_{3}\)  with \({x}_{2}+{x}_{3} \geq {y}_{2}+{y}_{3}\)} &
\multicolumn{1}{c}{The function \(f_{k}(i)\)} &
\multicolumn{1}{c}{The probabilities}\Tstrut\Bstrut \\ [1ex] 
\cline{1-1}  \cline{2-2}  \cline{3-3} \\ [-1ex]
  \({\pi}_{1}^{1}\ket{e_1} = \ket{e_5}\) ; \; \({\pi}_{2}^{1}\ket{e_2} = \ket{e_6}\) ; \;  
\({\pi}_{3}^{1}\ket{e_3} = \ket{e_7}\) ; \; \({\pi}_{4}^{1}\ket{e_4} = \ket{e_8}\)  & \(\varkappa_{11}; \; \varkappa_{22}; \; \varkappa_{33}; \; \varkappa_{44}\) &  
\(\mathfrak{p}_1 = 1-\sum_{k=2}^{4} \mathfrak{p}_k\)  \\   [2ex]
  \({\pi}_{1}^{2}\ket{e_1} = \ket{e_8}\) ; \; \({\pi}_{2}^{2}\ket{e_2} = \ket{e_6}\) ; \; 
\({\pi}_{3}^{2}\ket{e_3} = \ket{e_7}\) ; \; \({\pi}_{4}^{2}\ket{e_4} = \ket{e_5}\) & \(\varkappa_{41}; \; \varkappa_{22}; \; \varkappa_{33}; \; \varkappa_{14}\)  &
\(\mathfrak{p}_2={({x}_{4}-{y}_{4})}/{({y}_{1} - {y}_{4})}\)  \\   [2ex]
\({\pi}_{1}^{3}\ket{e_1} = \ket{e_5}\) ; \; \({\pi}_{2}^{3}\ket{e_2} = \ket{e_7}\) ; \; 
\({\pi}_{3}^{3}\ket{e_3} = \ket{e_6}\) ; \; \({\pi}_{4}^{3}\ket{e_4} = \ket{e_8}\)  & \(\varkappa_{11}; \; \varkappa_{32}; \; \varkappa_{23}; \; \varkappa_{44}\) &  
\(\mathfrak{p}_3={({y}_{2} - {x}_{2})}/{({y}_{2} - {y}_{3})}\)  \\   [2ex]
\({\pi}_{1}^{4}\ket{e_1} = \ket{e_7}\) ; \; \({\pi}_{2}^{4}\ket{e_2} = \ket{e_6}\) ; \; 
\({\pi}_{3}^{4}\ket{e_3} = \ket{e_5}\) ; \; \({\pi}_{4}^{4}\ket{e_4} = \ket{e_8}\) & \(\varkappa_{31}; \; \varkappa_{22}; \; \varkappa_{13}; \; \varkappa_{44}\) & \(\mathfrak{p}_4={({x}_{2} + {x}_{3} - {y}_{2} - {y}_{3})}/{({y}_{1} - {y}_{3})}\)  \\   [2ex]
\end{tabular}
\end{ruledtabular}
\end{table*}

\begin{table*}[htb!] 
\caption{The table includes the permutations \(\{{\pi}_{i}^{k}\}\), the function \(f_{k}(i) = \sqrt{{y_a}/{x_b}}\) denoted as \(\varkappa_{ab}\) for \(a, b = 1, 2, 3, 4\), and the corresponding probabilities \(\{\mathfrak{p}_k\}\). These integral elements contribute to the formulation of the local operators \(\{\hat{\mathcal{E}}_{A'}^{k}\}\) and \(\{\hat{\mathcal{E}}_{B'}^{k}\}\) given by Eqs.~\eqref{Local-Op-Epsilon-A} and \eqref{Local-Op-Epsilon-B}, respectively. In addition, the formulation of the local operators \(\{\hat{\mathcal{O}}_{A'}^{k}\}\) and \(\{\hat{\mathcal{O}}_{B'}^{k}\}\) given by Eqs.~\eqref{Local-Op-O-A} and \eqref{Local-Op-O-B}, respectively, follows a similar construction, where the substitution of variables is applied. Specifically, replacing \(\{{\pi}_{i}^{k}\}\) with \(\{{\pi}_{j}^{k}\}\), \(f_k(i)\) with \(g_k(j)\), and \(\mathfrak{p}\) with \(\mathfrak{q}\) yields the corresponding elements. Importantly, the entries in the table are applicable to the scenario where \({x}_{2} \leq {y}_{2}\) and \({x}_{3} \geq {y}_{3}\) with \({x}_{2}+{x}_{3} \leq {y}_{2}+{y}_{3}\). We have omitted the subscripts \(A'\) and \(B'\) in the representation of states for the sake of simplicity.}
\label{TablePermutations4D4th2}
\centering
\begin{ruledtabular}
\begin{tabular}{c c c}
\multicolumn{1}{c}{Permutations for the case \({x}_{2} \leq {y}_{2}\) \(\&\) \({x}_{3} \geq {y}_{3}\)  with \({x}_{2}+{x}_{3} \leq {y}_{2}+{y}_{3}\)} &
\multicolumn{1}{c}{The function \(f_{k}(i)\)} &
\multicolumn{1}{c}{The probabilities}\Tstrut\Bstrut \\ [1ex] 
\cline{1-1}  \cline{2-2}  \cline{3-3} \\ [-1ex]
  \({\pi}_{1}^{1}\ket{e_1} = \ket{e_5}\) ; \; \({\pi}_{2}^{1}\ket{e_2} = \ket{e_6}\) ; \;  
\({\pi}_{3}^{1}\ket{e_3} = \ket{e_7}\) ; \; \({\pi}_{4}^{1}\ket{e_4} = \ket{e_8}\) & \(\varkappa_{11}; \; \varkappa_{22}; \; \varkappa_{33}; \; \varkappa_{44}\)  &  
\(\mathfrak{p}_1 = 1-\sum_{k=2}^{4} \mathfrak{p}_k\)  \\   [2ex]
 \({\pi}_{1}^{2}\ket{e_1} = \ket{e_8}\) ; \; \({\pi}_{2}^{2}\ket{e_2} = \ket{e_6}\) ; \; 
\({\pi}_{3}^{2}\ket{e_3} = \ket{e_7}\) ; \; \({\pi}_{4}^{2}\ket{e_4} = \ket{e_5}\) & \(\varkappa_{41}; \; \varkappa_{22}; \; \varkappa_{33}; \; \varkappa_{14}\)  &
\(\mathfrak{p}_2={({y}_{1}-{x}_{1})}/{({y}_{1} - {y}_{4})}\)  \\   [2ex]
 \({\pi}_{1}^{3}\ket{e_1} = \ket{e_5}\) ; \; \({\pi}_{2}^{3}\ket{e_2} = \ket{e_7}\) ; \; 
\({\pi}_{3}^{3}\ket{e_3} = \ket{e_6}\) ; \; \({\pi}_{4}^{3}\ket{e_4} = \ket{e_8}\)  & \(\varkappa_{11}; \; \varkappa_{32}; \; \varkappa_{23}; \; \varkappa_{44}\) &  
\(\mathfrak{p}_3={({x}_{3} - {y}_{3})}/{({y}_{2} - {y}_{3})}\)  \\   [2ex]
  \({\pi}_{1}^{4}\ket{e_1} = \ket{e_5}\) ; \; \({\pi}_{2}^{4}\ket{e_2} = \ket{e_8}\) ; \; 
\({\pi}_{3}^{4}\ket{e_3} = \ket{e_7}\) ; \; \({\pi}_{4}^{4}\ket{e_4} = \ket{e_6}\) & \(\varkappa_{11}; \; \varkappa_{42}; \; \varkappa_{33}; \; \varkappa_{24}\) & \(\mathfrak{p}_4={({y}_{2} + {y}_{3} - {x}_{2} - {x}_{3})}/{({y}_{2} - {y}_{4})}\)  \\   [2ex]
\end{tabular}
\end{ruledtabular}
\end{table*}

\end{document}